\documentclass{aastex}
\usepackage{emulateapj5}
\usepackage{epsfig}
\usepackage{graphicx}

\def\ltsima{$\; \buildrel < \over \sim \;$}
\def\lsim{\lower.5ex\hbox{\ltsima}}
\def\gtsima{$\; \buildrel > \over \sim \;$}
\def\gsim{\lower.5ex\hbox{\gtsima}}

\shorttitle{CANDELS photo-z investigation}
\shortauthors{Dahlen et al.}

\begin{document}

%% LaTeX will automatically break titles if they run longer than
%% one line. However, you may use \\ to force a line break if
%% you desire.

%%\title{CANDELS Investigation of Photometric Redshift Methods}
\title{A Critical Assessment of Photometric Redshift Methods: A CANDELS Investigation}
%% Use \author, \affil, and the \and command to format
%% author and affiliation information.
%% Note that \email has replaced the old \authoremail command
%% from AASTeX v4.0. You can use \email to mark an email address
%% anywhere in the paper, not just in the front matter.
%% As in the title, you can use \\ to force line breaks.

\author{Tomas Dahlen\altaffilmark{1}, 
Bahram Mobasher\altaffilmark{2},
Sandra M. Faber\altaffilmark{3},
Henry C Ferguson\altaffilmark{1},
Guillermo Barro \altaffilmark{3},
Steven L. Finkelstein \altaffilmark{4},
Kristian Finlator \altaffilmark{5},
Adriano Fontana \altaffilmark{6},
Ruth Gruetzbauch \altaffilmark{7},
Seth Johnson  \altaffilmark{8},
Janine Pforr  \altaffilmark{9},
Mara Salvato  \altaffilmark{10,11},
Tommy Wiklind  \altaffilmark{12},
Stijn Wuyts  \altaffilmark{10},
Viviana Acquaviva\altaffilmark{13}, 
Mark E. Dickinson \altaffilmark{9},
Yicheng Guo  \altaffilmark{3},
Jiasheng Huang \altaffilmark{14,15},
Kuang-Han Huang \altaffilmark{16},
Jeffrey A. Newman \altaffilmark{17},
Eric F. Bell \altaffilmark{18},
Christopher J. Conselice \altaffilmark{19},
Audrey Galametz \altaffilmark{6},
Eric Gawiser  \altaffilmark{20},
Mauro Giavalisco  \altaffilmark{8},
Norman A. Grogin  \altaffilmark{1},
Nimish Hathi \altaffilmark{21},
Dale Kocevski \altaffilmark{22},
Anton M. Koekemoer \altaffilmark{1},
David C. Koo  \altaffilmark{3},
Kyoung-Soo Lee \altaffilmark{23},
Elizabeth J. McGrath \altaffilmark{24},
Casey Papovich \altaffilmark{25},
Michael Peth \altaffilmark{16},
Russell Ryan  \altaffilmark{1},
Rachel Somerville  \altaffilmark{20},
Benjamin Weiner  \altaffilmark{26},
and Grant Wilson  \altaffilmark{8}
}
\email{dahlen@stsci.edu}

\altaffiltext{1}{Space Telescope Science Institute, 3700 San Martin Drive, Baltimore, MD 21218}
\altaffiltext{2}{Department of Physics and Astronomy, University of California, Riverside, CA 92521}
\altaffiltext{3}{UCO/Lick Observatory, Department of Astronomy and Astrophysics, University of California, Santa Cruz, CA 95064}
\altaffiltext{4}{Department of Astronomy, The University of Texas at Austin, Austin, TX 78712}
\altaffiltext{5}{Dark Cosmology Centre, Niels Bohr Institute, University of Copenhagen, Denmark}
\altaffiltext{6}{INAF - Osservatorio Astronomico di Roma, Via Frascati 33, I–00040, Monteporzio, Italy}
\altaffiltext{7}{Center for Astronomy and Astrophysics, Observatorio Astronomico de Lisboa, Tapada da Ajuda, 1349-018 Lisboa, Portugal}
\altaffiltext{8}{Department of Astronomy, University of Massachusetts, 710 North Pleasant Street, Amherst, MA 01003}
\altaffiltext{9}{NOAO, 950 N. Cherry Avenue, Tucson, AZ 85719}
\altaffiltext{10}{Max-Planck-Institut f\"ur extraterrestrische Physik, Giessenbachstrasse 1, D-85748 Garching bei M\"{u}nchen, Germany}
\altaffiltext{11}{Excellence Cluster, Boltzmann Strasse 2 D-85748, Garching, Germany}
\altaffiltext{12}{Joint ALMA Observatory, Alonso de Cordova 3107, Vitacura, Santiago, Chile}
\altaffiltext{13}{Physics Department, CUNY NYC College of Technology, 300 Jay Street, Brooklyn, NY 11201}
\altaffiltext{14}{Harvard-Smithsonian Center for Astrophysics, 60 Garden Street, Cambridge, MA 02138}
\altaffiltext{15}{National Astronomical Observatories, Chinese Academy of Sciences, Beijing 100012, China}
\altaffiltext{16}{Department of Physics and Astronomy, Johns Hopkins University, 3400 North Charles Street, Baltimore, MD 21218}
\altaffiltext{17}{Department of Physics and Astronomy, University of Pittsburgh, Pittsburgh, PA 15260}
\altaffiltext{18}{Department of Astronomy, University of Michigan, 500 Church Street, Ann Arbor, MI 48109}
\altaffiltext{19}{School of Physics and Astronomy, University of Nottingham, Nottingham, UK}
\altaffiltext{20}{Department of Physics and Astronomy, Rutgers, The State University of New Jersey, 136 Frelinghuysen Road, Piscataway, NJ 08854}
\altaffiltext{21}{Carnegie Observatories, 813 Santa Barbara Street, Pasadena, CA 91101}
\altaffiltext{22}{Department of Physics and Astronomy, University of Kentucky, Lexington, KY 40506}
\altaffiltext{23}{Department of Physics, Purdue University, 525 Northwestern Avenue, West Lafayette, IN 47907}
\altaffiltext{24}{Department of Physics and Astronomy, Colby College, Waterville, ME 04901}
\altaffiltext{25}{Department of Physics and Astronomy, Texas A\&M University, College Station, TX 77843}
\altaffiltext{26}{Steward Observatory, 933 North Cherry Street, University of Arizona, Tucson, AZ 85721}

\begin{abstract}
We present results from the Cosmic Assembly Near-infrared Deep Extragalactic Legacy Survey (CANDELS) photometric redshift methods investigation. In this investigation, the results from eleven participants, each using a different combination of photometric redshift code, template spectral energy distributions (SEDs) and priors, are used to examine the properties of photometric redshifts applied to deep fields with broad-band multi-wavelength coverage. The photometry used includes $U$-band through mid-infrared filters and was derived using the TFIT method. Comparing the results, we find that there is no particular code or set of template SEDs that results in significantly better photometric redshifts compared to others. However, we find codes producing the lowest scatter and outlier fraction utilize a training sample to optimize photometric redshifts by adding zero-point offsets, template adjusting or adding  extra smoothing errors. These results therefore stress the importance of the training procedure. We find a strong dependence of the photometric redshift accuracy on the signal-to-noise ratio of the photometry. On the other hand, we find a weak dependence of the photometric redshift scatter with redshift and galaxy color. We find that most photometric redshift codes quote redshift errors (e.g., 68\% confidence intervals) that are too small compared to that expected from the spectroscopic control sample. We find that all codes show a statistically significant bias in the photometric redshifts. However, the bias is in all cases smaller than the scatter, the latter therefore dominates the errors. Finally, we find that combining results from multiple codes significantly decreases the photometric redshift scatter and outlier fraction. We discuss different ways of combining data to produce accurate photometric redshifts and error estimates.
\end{abstract}


\section{Introduction}
Using photometric redshifts to estimate the distances of faint galaxies has become an  integral part of galaxy surveys conducted during recent years. This is driven by the large number of galaxies, and their faint fluxes which have made spectroscopic follow-up infeasible except for a relatively small and bright fraction of the galaxy  population. Albeit less precise and less accurate than spectroscopy, photometric redshifts provide a way to estimate distances for galaxies too faint for spectroscopy or samples too large to be practical for complete spectroscopic coverage. Since the early description of using colors to determine distances in Baum (1962), and the important developments over the years described in e.g., Koo (1985), Connolly et al. (1995) and Gwyn (1995), the number of articles describing the method and the number of applications for photometric redshifts have grown rapidly.

The photometric redshift technique is usually divided into two groups, template fitting and empirical fitting. The template fitting technique derives the photometric redshift by minimizing the value $\chi^2$~when comparing an observed SED with the SED computed from a template library that includes spectral-energy distributions for a variety of galaxy types (representing different redshifts, star-formation histories, chemical abundance, and mixtures of dust and stars). The empirical technique uses a training set of galaxies with known spectroscopic redshifts to derive a relation between observed photometry and redshifts. Today, a large number of codes of both techniques exists, many of which are publicly available. Codes based on the template fitting technique include: zphot (Giallongo et al. 1998), HyperZ (Bolzonella et al. 2000), BPZ (Ben\'{i}tez 2000), ImpZ (Babbedge et al. 2004), ZEBRA (Feldmann et al. 2006), SPOC (Finlator et al. 2007), EAZY (Brammer et al. 2008), Low Resolution Template (LRT) Libraries (Assef et al. 2008), GALEV (Kotulla et al. 2009), Rainbow (Barro et al. 2011), GOODZ (Dahlen et al. 2010), LePhare (Ilbert et al. 2006; S. Arnouts \& O. Ilbert 2013, in preparation), and SATMC (S. Johnson et al. 2013, in preparation). Empirical codes include: ANNz (Collister \& Lahav 2004); Multilayer Perceptron Artificial Neural Network (Vanzella et al. 2004); ArborZ (Gerdes et al. 2010); ``Empirical-$\chi^2$'' (Wolf 2009); ``Random Forests'' (Carliles et al. 2010). Certain codes combine the methodology of both techniques (e.g., EAZY, GOODZ, and LePhare) which can use a training set of galaxies to derive corrections to zero-points and/or template SED shapes in order to minimize the scatter between photometric and spectroscopic redshifts in the training sample. These corrections can then be applied to the full set of galaxies without spectroscopy.

The Cosmic Assembly Near-infrared Deep Extragalactic Legacy Survey (CANDELS; PIs S. Faber and H. Ferguson; see Grogin et al. 2011 and Koekemoer et al. 2011) is an $HST$~Multi-Cycle Treasury program aimed at imaging distant galaxies in multiple wavebands and detect high redshift supernovae in five sky regions: the GOODS-S, GOODS-N, EGS, UDS, and COSMOS fields. Images and catalogs will be provided to the public for the different fields. Besides photometry, the catalogs will include auxiliary information such as photometric redshifts and stellar masses of galaxies. The CANDELS data include some of the deepest photometry available in both optical and infrared over a wide area and it is important to investigate the behavior of the derived quantities at the faint flux levels typical of the survey. Therefore, the CANDELS team has preformed a series of tests to evaluate how photometric redshift and mass estimates from different codes compare, how well codes reproduce the redshift of objects with spectroscopic redshifts, how well codes reproduce masses from simulated galaxies, and how photometric redshift estimates depend on signal-to-noise, redshift and galaxy color. Furthermore, we investigate how the error estimates determined by the codes compare with the errors expected from either spectroscopic control samples or simulated galaxy catalogs. Finally, we investigate possible ways of combining results from individual codes in order to improve the quality of the photometric redshifts.

While the investigation was performed with the CANDELS data in mind, the questions should be general and the results relevant for any survey targeting distant galaxies. In this paper, we focus the investigation on the photometric redshift technique. A number of collaborators in the CANDELS team were asked to use their preferred photometric redshift code to derive redshifts for a set of photometric catalogs. The results from the different codes and sets of template SEDs were thereafter compared with the aim of deriving the best photometric redshifts possible given the available data set and to minimize possible biases in the derived redshifts. In an accompanying paper, B. Mobasher et al . (2013, in preparation), we discuss estimates of stellar masses using the same catalogs.

This paper is organized as follows: In Section 2, we describe the catalogs used in the testing followed in Section 3 by a presentation of the different codes used. Results are given in Section 4, followed by a discussion on ways to combine data to improve photometric redshifts in Section 5. Section 6 presents a comparison to earlier work. A summary is given in Section 7. Throughout we assume a cosmology with $\Omega_M$=0.3, $\Omega_{\Lambda}$=0.7, and h=0.7. Magnitudes are given in the AB system.

\section{Test Catalogs}
Two different catalogs were used to test the photometric redshifts. The first is a near-IR $HST$/WFC3 $H$-band (F160W filter) selected catalog, while the second is an optical $HST$/ACS $z$-band (F850LP filter) selected catalog. Both catalogs cover the GOODS-S area (Giavalisco et al. 2004), with photometry derived using the TFIT method (Laidler et al. 2007). We use two fairly similar catalogs, to investigate possible differences in optical versus near-IR selected photometric redshifts. For both catalogs we provided a test sample with known spectroscopic redshifts for training the photometric redshift codes. Each participant in the CANDELS SED-fitting test was asked to derive photometric redshifts for the objects in each catalog, including both the training sample and a control sample for which the redshifts were not provided.  Below we give more details on the different catalogs.

\subsection{WFC3 $H$-band selected catalog}
The primary test catalog includes the $HST$/WFC3 $H$-band selected TFIT multi-band photometry. The catalog contains 20,000 objects in the GOODS-S field and includes photometry in 14 bands:  $U$~(VLT/VIMOS), $BViz$~($HST$/ACS), F098M, F105W, F125W, F160W (WFC3/IR), $K_s$~(VLT/ISAAC) and 3.6, 4.5, 5.8, 8.0 micron ($Spitzer$/IRAC). The total area covered in the catalog is approximately $\sim$100 arcmin$^2$. Note that F098M covers $\sim$40\% of the area (data taken from the Early Release Science, Windhorst et al. 2011), while F105W covers most of the remaining $\sim$60\%, therefore, 13 band photometry is the maximum for any individual object. Photometry in the ACS and WFC3 bands are measured using SExtractor in dual-image mode with F160W as the detection band. For all other bands, the TFIT method was used. This results in a flux measurement for all objects in all bands that cover the footprint of the F160W data. Both SExtractor and TFIT will provide flux estimates for sources based on prior information on position and shape from the H-band image. Therefore fluxes are provided in every band even for sources that are not formally detected in that band. These fluxes can sometimes be negative due to statistical fluctuations. If the photometric error estimates are corrected, this should not cause problems for the photometric-redshift estimates. We also note that the F160W band photometry includes 4 of the 10 planned epochs of GOODS-S data available at the time of the test. A detailed description of the CANDLES GOODS-S data is given in Guo et al. (2013). The methodology used to derive the photometry is described in Galametz et al. (2013).

\subsection{ACS z-band selected catalog}
As a secondary test catalog, we use an ACS $z$-band selected TFIT catalog of GOODS-S that includes multi-waveband photometry in twelve bands: $U$~(VLT/VIMOS), $BViz$~($HST$/ACS), $JHKs$~(VLT/ISAAC) and 3.6, 4.5, 5.8, 8.0 micron ($Spitzer$/IRAC).  The data is the same as for the primary test catalog except that ISAAC $J$~and $H$~are added and WFC3 IR bands are excluded. The area covered by the $z$-band selected catalog is $\sim$150 arcmin$^2$~and the number of objects included in the catalog is 25,000. We use the secondary catalog to examine the effect of selecting the catalog in the optical vs. near-IR WFC3 when estimating photometric redshifts. Details on the photometry are given in Dahlen et al. (2010). 

\subsection{Spectroscopic comparison sample}
We use a sample of galaxies with known spectroscopic redshifts to evaluate how well the photometric redshifts reproduce the true redshifts as given by the spectra. Our spectroscopic sample is compiled from a set of publicly available data including Cristiani et al. (2000), Croom et al. (2001), Bunker et al. (2003), Dickinson et al. (2004), LeF\`{e}vre et al. (2004), Stanway et al. (2004), Strolger et al. (2004), Szokoly et al. (2004), van der Wel et al. (2004), Doherty et al. (2005), Mignoli et al. (2005), Roche et al. (2006), Ravikumar et al. (2007), Vanzella et al. (2008), and Popesso et al. (2009).

\begin{figure*}
\epsscale{1.0}
\plotone{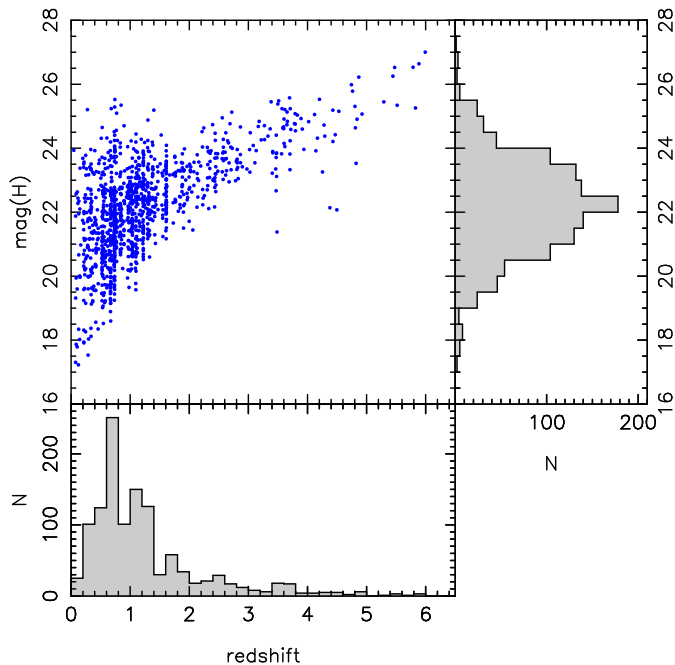}
\figcaption[f1.eps]{Redshift and $H$-band magnitude distributions of the spectroscopic sample used to train and evaluate the photometric redshifts.
\label{fig1}}
\end{figure*}

When selecting the sources for inclusion in our spectroscopic redshift sample, we specifically include only objects with the highest possible data quality (when available). Furthermore, we exclude all objects with X-ray detection in the Chandra 4Ms sample from Xue et al. (2011) and radio sources in Afonso et al. (2006) and Padovani et al. (2011). Even though there are more than 3000 spectroscopic redshifts in the GOODS-S ACS footprint, we exclude more than half of these to minimize the number of faulty redshifts and AGN contaminants. The latter are excluded since the aim here is to derive and compare photometric redshifts for a population of ``normal'' galaxies. Photometric redshifts for X-ray sources are discussed in Salvato et al. (2009, 2011, 2013 in preparation). We divide the final set of highest quality spectroscopic redshifts into a training sample provided to each participant in the test. A second control sample is used to evaluate the accuracy of the photometric redshifts. Both catalogs cover the same ranges in magnitude, color, and redshift. The training catalogs include 580 and 640 objects, while the control samples contain 589 and 614 objects for the $H$-selected and $z$-selected catalogs, respectively. The difference in the total number of objects between the different selections is due to the difference in covered area. The redshift and magnitude distributions of the spectroscopic sample are presented in Figure \ref{fig1}.

\subsection{Publicly available test catalogs}
The GOODS-S $H$-band selected test catalogs and associated files are available via the STScI Archive High-Level Science Products page for CANDELS\footnote{\tt http://archive.stsci.edu/prepds/candels}. This includes the 14 band photometry and spectroscopic redshifts for 580 and 589 objects in the training and control samples, respectively.

\section{Participating codes}
A total of thirteen submissions to the CANDELS SED-fitting test were received and each participant was given an ID number. Of these thirteen, eleven included calculated photometric redshifts, while the remaining two only presented derived masses (for objects with known spectroscopic redshifts).  In Table \ref{table1}, we list the eleven participants that provided photometric redshifts (participants only producing masses are described in B. Mobasher et al. 2013, in preparation) and the name of the photometric redshift code used. Each different code is assigned a single character code identifier in the range A-I. We hereafter refer the to the combination of participant and photometric redshift code by combining the two identifiers, e.g., 2A, 3B, 4C and so on. This makes it easy to identify the participants that use the same photometric redshift code, i.e., 4C, 7C, and 13C. For simplicity, we refer to the eleven different participant and code combinations as ``codes'' in the following. The table lists the template set used and shows if emission lines are included. We provide the latter information since emission lines can have a significant effect on broad-band photometry and therefore the template SED fitting (e.g., Atek et al. 2011; Schaerer \& de Barros 2012; Stark et al. 2013). Also shown is if the codes use the control sample of spectroscopic redshifts to calculate a ``flux shift'' to the given photometry or template SEDs. Indicated is also if the code adds an extra error to the provided flux errors when template fitting. The most common way of implementing additional errors is to add (in quadrature) an error corresponding to 2-10\% of the flux ($\sim$0.02-0.1 mag) to the given photometric errors. Alternatively, a lower limit to the given errors can be enforced. Finally, the table indicates if the code adjusts the template SEDs based on the training sample and if the code uses interpolations between template SEDs.
\begin{table*}
\caption{Codes included in the CANDELS SED test for calculating photometric redshifts.}
\centering
\begin{tabular}{rllclcccccc}
ID$^a$ & PI & Code & Code ID & Template set & Em lines & Flux shift & $\Delta$err & $\Delta$SED & Inter & ref. \\
\tableline 
\tableline
2 & G. Barro & Rainbow & A &  PEGASE$^b$ & yes& yes & no & no & no  & $j$\\
3 & T. Dahlen & GOODZ & B &  CWW$^c$, Kinney$^d$ &yes & yes & yes & yes & yes & $k$\\
4 & S. Finkelstein & EAZY & C &  EAZY$^e$+BX418$^f$ &yes& no & no & no & yes & $l$\\
5 & K. Finlator & SPOC & D &  BC03$^g$ &yes& no & no & no & no  & $m$ \\
6 & A. Fontana & zphot & E &  PEGASEv2.0$^b$ &yes& yes & yes & no & no  & $n,o$\\
7 & R. Gruetzbauch & EAZY & C &  EAZY$^e$ &yes& yes & yes & no & yes  & $l$\\
8 & S. Johnson & SATMC & F & BC03$^g$ &no& no & no & no & yes  & $p$\\
9 & J. Pforr & HyperZ & G & Maraston05$^h$ &no& no & yes & no & no  & $q$ \\
11 & M. Salvato & LePhare & H &  BC03$^g$+Polletta07$^i$ &yes& yes & yes & no & no & $r$ \\
12 & T. Wikind & WikZ & I & BC03$^g$ &no& no &  yes & no & no & $s$\\
13 & S. Wuyts & EAZY &  C & EAZY$^e$ &yes& yes & yes & no & yes   & $l$\\
\tableline
\end{tabular}
\tablecomments{
Col 1: ID number of participant. Col 2: Name of photometric redshift investigator. Col 3: Name of code. Col 4: Code identifier. Col 5: Template SED used to derive photometric redshifts. Col 6: Are emission lines included in template SEDs (yes/no). Col 7: Applies shifts to the fluxes or templates based on spectroscopic training sample (yes/no) Col 8: Adds extra errors to the fluxes in addition to fluxes given in the photometric catalogs (yes/no). Col 9: Adjusts template SEDs based on spectroscopic training set (yes/no). Col 10: Uses interpolations between template SEDs.  Col 11: Reference to code.\\
$^{a}$ Codes which ID 1 and 10 are not used to calculate photometric redshift in this test, however they are used to calculate masses in the accompanying paper by B. Mobasher et al. (2013, in preparation), 
%$^{1}$ S. Charlot \& G. Bruzual (2012, in prep.), 
$^b$ Fioc \& Rocca-Volmerange (1997), 
$^c$ Coleman et al. (1980), 
$^d$ Kinney et al. (1996), 
$^e$ The EASY template set from Brammer et al. (2008) consists of six templates based on the PEGASE models (Fioc \& Rocca-Volmerange 1997), 
$^f$ Erb et al. (2010), 
$^g$ Bruzual \& Charlot (2003), 
$^h$ Maraston (2005), 
$^i$ Polletta et al. (2007),
$^j$ Barro et al. (2011),
$^k$ Dahlen et al. (2010),
$^l$ Brammer et al. (2008),
$^m$ Finlator et al. (2007),
$^n$ Giallongo et al. (1998),
$^o$ Fontana et al. (2000),
$^p$ S. Johnson et al. (2013, in prep.),
$^q$ Bolzonella et al. (2000),
$^r$ S. Arnouts \& O. Ilbert (2013, in prep.), and
$^s$ Wiklind et al. (2008).
}
\label{table1}
\end{table*}
Below we give a short summary of each code participating in the photometric redshift test. For each code we describe if the $\chi^2$~minimization is done in magnitude or flux space and how negative fluxes are treated. We also note the codes using priors in the fitting. We finally comment on any special treatment of the IRAC fluxes in the fitting, such as excluding the 5.8$\mu$m and 8.0$\mu$m channels (hereafter ch3 and ch4) at low redshift where they probe wavelengths where templates may not be as reliable. For details, please see the quoted articles.\\
\\
A - {\em Rainbow} (P\'{e}rez-Gonz\'{a}lez et al. 2008; Barro et al. 2011)\footnote{\tt https://rainbowx.fis.ucm.es/Rainbow\_Database/}\\
A template fitting code based on $\chi^2$~minimization between observed photometry and a set of $\sim$1500 semi-empirical template SEDs computed from spectroscopically confirmed galaxies modeled with PEGASE stellar population synthesis models (see P\'{e}rez-Gonz\'{a}lez et al. 2008, Appendix A). The code allows for additional smoothing errors, photometric zero-point corrections and a template error function to down-weight the wavelength ranges where the templates a more uncertain (e.g., the rest frame near-IR). Fitting is done in flux space. Negative fluxes and data points with signal-to-noise $<$2.5 are not included in the fitting. Excludes IRAC bands that probe rest frame $> 5\mu m$~(ch3 at $z\lsim$ 0.15 and ch4 at $z\lsim$ 0.6). Run by G. Barro.\\
\\
B - {\em GOODZ} (Dahlen et al. 2010)\\
A template fitting code that minimizes $\chi^2$~between observed photometry and a set of template SEDs. The code allows the use of luminosity function priors. For this test, a rest frame $V$-band luminosity function prior was used. This assignes a low probability at low redshifts where the volume element is small and a low probability for bright objects at high redshifts. The code also calculates and applies shifts to the input photometry based on a training set of galaxies. Can adjust templates using a training sample. Extra smoothing errors were added to existing photometric errors.  Includes the option of using interpolations between the provided template SEDs. Fitting is done in flux space. Negative fluxes are included in the fitting. Excludes IRAC bands that probe rest frame $> 5\mu m$~(ch3 at $z\lsim$ 0.15 and ch4 at $z\lsim$ 0.6). Run by T. Dahlen.\\ 
\\
C - {\em EAZY} (Brammer et al. 2008)\footnote{\tt http://www.astro.yale.edu/eazy/}\\
Also a template fitting code based on $\chi^2$~minimization between observed photometry and template SEDs. Magnitude priors can be included and the code can apply shifts to the input photometry. Extra smoothing errors can be added to existing errors in the photometric catalogs. Includes the option of using interpolations between the provided template SEDs. Fitting is done in flux space. Run by S. Finkelstein, R. Gruetzbauch, and S. Wuyts. When doing the fitting, SF and SW include negative fluxes, while RG ignores those data points. EAZY includes the option to apply a template error function that down-weights datapoints at rest frame $\gsim$2$\mu$m. This option is used by all three participants running this code. A luminosity function prior can also be included. This assignes a low probability where the volume sampled is small (low redshifts) and a low probability at high redshifts for objects with bright apparent magnitudes. A K-band luminosity function prior was used by SW.\\ 
\\
D - {\em SPOC} (Finlator et al. 2007)\\
A $\chi^2$~minimization code, however the template SEDs are derived directly from cosmological numerical simulations. Numerically simulated SFHs and metallicities are directly adopted, rather than assuming toy-model SFHs (for example, constant or declining), leaving effectively three free parameters, namely M*, $z$, and A$_V$. However, the code is unlikely to find a match for any galaxy whose stellar mass lies below the mass resolution limit of the simulations from which the template library was extracted (since the code does not scale the luminosity of the galaxies independently). For the  version of the code used in this investigation, the mass limit is ~1.4$\times10^9M/M_{\odot}$, resulting in matches for only about half of the objects in the spectroscopic sample. Fitting is performed in flux space. Objects with negative fluxes are treated as though the negative flux indicated -1$\times$~the 1$\sigma$ upper limit (i.e., the flux error).  In this case, model SEDs that are brighter than 3$\times$~that 1$\sigma$~upper limit in that band are rejected outright, otherwise the band does not contribute to the $\chi^2$. Run by K. Finlator.\\
\\
E - {\em zphot} (Giallongo et al. 1998; Fontana et al. 2000)\\
A $\chi^2$ minimization code using template SEDs. It is flexible, allowing the user to adopt a wide variety of templates, including synthetic models taken from  BC03, Maraston (2005), and Fioc \& Rocca-Volmerange et al. (1997), with various choices of the Star Formation History, as well as observed templates of stars, galaxies and AGNs from a variety of sources. 
Dust extinction and IGM absorption can also be added. For the redshift determination in this investigation, a library composed of templates taken from the Fioc \& Rocca-Volmerange library has been used, with the addition of Calzetti et al. (2000) extinction and Fan et al. (2006) IGM absorption.
A minimum photometric errors can be set in each band. zphot can accept both fluxes and magnitudes in the input catalog, and computes the photometric redshifts directly from the best-matching template. The code also computes the error estimate on the fitted values by scanning the probability distribution of the various parameters. For this test, fitting is done in flux space including negative fluxes. When the flux is $<$-flux-error, the flux is set to zero in order to prevent datapoints with unrealistically negative fluxes to inflate the $\chi^2$~value. Excludes IRAC bands that probe rest frame $> 5\mu m$~(ch3 at $z\lsim$ 0.15 and ch4 at $z\lsim$ 0.6). Run by A. Fontana. \\
\\
F - {\em SATMC} (S. Johnson et al. 2013, in prep)\\
A general purpose SED fitting routine using Monte Carlo Markov Chain (MCMC) techniques. The SED fits are performed in likelihood space, which are computed in a
similar manner to standard $\chi^2$~techniques.  During the fits, all available parameters are allowed to vary. Utilizing the BC03 template set, these include galaxy age, E(B-V) extinction, e-folding timescale for the SFH in addition to the photometric redshift and normalization (i.e. stellar mass).  Fitting is done in flux space. For negative fluxes, an upper limit (basing the upper limits on 3 times the flux errors) method which follows a one-sided Gaussian distribution is used. Effectively, models with fluxes below the upper limit are always accepted while those with higher flux values are given a proportionally small probability during the fits. Run by S. Johnson. \\
\\
G - {\em HyperZ} (Bolzonella et al. 2000)\footnote{\tt http://webast.ast.obs-mip.fr/hyperz/}\\
This is a $\chi^2$~minimization code. Shifts to magnitudes can be added manually. A minimum photometric error can be set, errors smaller than this value will be replaced by the minimum value. Fitting is done in flux space. Negative fluxes are not included in the fitting. Uses a prior that requires the NIR absolute magnitude to be in the range $-30<M<-9$ (Vega mag). Run by J. Pforr.\\
\\
H - {\em LePhare} (S. Arnouts \& O. Ilbert 2013, in prep.)\footnote{\tt http://www.cfht.hawaii.edu/~arnouts/LEPHARE/lephare.html} \\
Another $\chi^2$~template SED fitting code, which can use a training sample to derive zero-point offsets and to optimize the template SEDs. The code also has the option of using luminosity priors and the possibility of adding extra errors to the given photometry. The code can output both the photometric redshift based on the minimum $\chi^2$~and the median of the photometric redshift probability distribution. The code can be run with or without emission line corrections, as described in Ilbert et al. (2006; 2009). Fitting can be done in either magnitude or flux space. Fitting during this investigation is done in magnitude space and negative fluxes are not included in the fitting. Uses a prior that requires the optical absolute magnitude to be in the range $-24<M<-8$. IRAC ch3 and ch4 are not used in the fitting. Run by. M. Salvato.\\
\\
I - {\em WikZ} (Wiklind et al. 2008) \\
A pure template fitting code, minimizing the $\chi^2$~between the observed and template SED photometry. The code has the possibility to add extra smoothing errors to the existing photometric errors. Fitting is done in flux space. For negative fluxes, the data point adds to the $\chi^2$~if the template SED flux is brighter than the 1$\sigma$ upper limit. If the template flux is lower than the upper limit, it does not add to the $\chi^2$. Does not include IRAC ch3 at $z<0.5$~and ch4 at $z<0.7$. Run by T. Wiklind.  \\
\\
Of the eleven submissions including photometric redshifts, nine different photometric redshift codes have been used. Only EAZY has been used by multiple participants, i.e., codes 4C, 7C, and 13C. However, there are some differences in the details of the template sets used in each case, e.g., code 4C includes a template of BX418 (Erb et al. 2010), a metal poor galaxy with strong Ly$\alpha$~emission, code 7C uses the EAZY templates, plus a template with deep Ly$\alpha$~absorption (constructed from an observed high-$z$~galaxy), while finally code 13C uses the six EASY templates after updating them by adding emission lines using the Ilbert et al. (2009) algorithm. Furthermore codes calculate and apply slightly different zero-point shifts and uses different smoothing to the existing photometric errors. In the $\chi^2$~fitting, code 4C and 13C include data points with negative fluxes while code 7C ignores them and only code 13C uses luminosity function priors.  Therefore, the codes are sufficiently different to produce independent estimates of the photometric redshifts. In the Section 4, we show that the scatter between the codes using EAZY is similar to the scatter between different codes.
\section{Results}
In Table \ref{table2} and Table \ref{table3}, we show the resulting scatter between the photometric redshifts and spectroscopic redshifts for the different codes presented in Table \ref{table1}. The scatter is calculated using the control sample only. The tables present the full scatter
\begin{equation}
\sigma_F=rms[\Delta z/(1+z_{spec})]
\end{equation}
where $\Delta$z=$z_{spec}-z_{phot}$.
Results are also given in $\sigma_O$, which is the rms after excluding outliers, where an outlier is defined as an object with $|\Delta z|/(1+z_{spec})>0.15$. Since many recent results in photometric redshift present scatter as the normalized median absolute deviation (Ilbert et al. 2009), we also give results as: 
\begin{equation} 
\sigma_{NMAD}=1.48\times median(\frac{|\Delta z|}{1+z_{spec}})
\end{equation}
Finally, we also calculate the scatter $\sigma_{dyn}$~using a dynamic definition of the outlier fraction. Here outliers are defined as objects with  $|\Delta z|/(1+z_{spec})>3\times\sigma_{dyn}$. The scatter and outlier fraction (OLF$_{dyn}$) are here determined iteratively. For a Gaussian distribution of the scatter, the outlier fraction would be constant ($\sim$0.3\%) regardless of the width of the distribution. However, since the scatter in the different codes are expected to be highly non Gaussian, the outlier fraction will vary between codes.

Furthermore, to quantify any systematic bias between photometric and spectroscopic redshifts, we define bias$_z$=mean[$\Delta z/(1+z_{spec})$], after excluding outliers (using the constant definition).
\begin{table*}
\caption{Photometric redshift results for WFC3 $H$-band selected catalog.}
\centering
\begin{tabular}{rcrccclcc}
Code & Objects & bias$_z^a$ & OLF$^b$ & $\sigma_F^c$ & $\sigma_O^d$ & $\sigma_{NMAD}^e$ &  $\sigma_{dyn}^f$ & OLF$_{dyn}^g$   \\
\tableline 
\tableline
2A &589 & -0.010 & 0.092 & 0.167 & 0.041 & 0.038     & 0.038& 0.107\\
3B & 589 &  -0.007 & 0.036 & 0.099 &  0.035 & 0.034 & 0.033& 0.048\\
4C & 589 & -0.009 & 0.051 & 0.114 & 0.044 & 0.040    &0.042 & 0.061\\
5D & 408 & -0.030 & 0.147 & 0.197 & 0.073 & 0.097   & 0.098 & 0.034\\
6E & 589 &  -0.007 & 0.041 & 0.104 & 0.037 & 0.033  & 0.033& 0.065\\
7C & 589 & -0.009 & 0.053 & 0.121 & 0.037 & 0.033   & 0.033& 0.070\\
8F & 589 & -0.008 & 0.093 & 0.272 & 0.064 & 0.077    & 0.074&0.051\\
9G & 589 &  0.013 & 0.078 & 0.189 & 0.050 & 0.045   & 0.053& 0.063\\
11H & 589 & -0.008 & 0.048 & 0.132 & 0.038 & 0.033 & 0.030&0.088\\
12I & 589 & -0.023 & 0.046 & 0.153 & 0.049 & 0.054 & 0.049&0.046\\
13C & 589 & -0.005 & 0.039 & 0.127 & 0.034 & 0.026 & 0.027&0.075\\
\hline
median(all) & 589 & -0.008& 0.029 & 0.088 & 0.031 & 0.029 & 0.026 & 0.054\\
median(5)  & 589 & -0.009& 0.031 & 0.079 & 0.029 & 0.025 & 0.024 & 0.056\\
\tableline
\end{tabular}
\tablecomments{
$^a$~bias$_z$=mean[$\Delta z/(1+z_{spec})$] after excluding outliers, where  $\Delta$z=$z_{spec}-z_{phot}$. 
$^b$OLF=Outlier fraction, i.e., fraction of objects that are outliers defined as $|\Delta z|/(1+z_{spec})>0.15$. 
$^c$~$\sigma_F=rms[\Delta z/(1+z_{spec})]$.
$^d$~$\sigma_O=rms[\Delta z/(1+z_{spec})]$~after excluding outliers.
$^e$~$\sigma_{NMAD}=1.48\times median(\frac{|\Delta z|}{1+z_{spec}})$.
$^f$~$\sigma_{dyn}$~rms after excluding outliers with $\Delta z/(1+z_{spec})>3\sigma_{dyn}$.
$^g$~ OLF$_{dyn}^g$~fraction outliers defined as objects with $\Delta z/(1+z_{spec})>3\sigma_{dyn}$.
The last two rows show the results after adopting the median photometric redshift of all codes, and the median of the five codes with overall lowest scatter, when calculating the scatter versus the spectroscopic sample.
}
\label{table2}
\end{table*}
\begin{table*}
\caption{Photometric redshift results for ACS $z$-band selected catalog.}
\centering
\begin{tabular}{rcrccclcc}
ID & Objects & bias$_z$ & OLF & $\sigma_F$ & $\sigma_O$ & $\sigma_{NMAD}$ &  $\sigma_{dyn}$ & OLF$_{dyn}$   \\
\tableline 
\tableline
2A & 614 & -0.018 & 0.086 & 0.259 & 0.052 & 0.054 & 0.053& 0.083\\
3B & 614 &  -0.004 & 0.057 & 0.148 &  0.039 & 0.034 & 0.032&0.091\\
4C & 614 & -0.011 & 0.077 & 0.197 & 0.046 & 0.045 & 0.045&0.083\\
5D & 446 & -0.032 & 0.067 & 0.259 & 0.070 & 0.087 & 0.080&0.029\\
6E & 614 &  -0.010 & 0.052 & 0.198 & 0.044 & 0.040 & 0.041&0.065\\
7C & 614 & -0.008 & 0.046 & 0.149 & 0.039 & 0.038 & 0.036&0.064\\
8F & 614 & -0.012 & 0.140 & 0.535 & 0.064 & 0.079 & 0.080&0.073\\
9G & 614 &  0.015 & 0.121 & 0.269 & 0.053 & 0.057 & 0.059&0.096\\
11H & 614 & -0.009 & 0.042 & 0.131 & 0.040 & 0.036 &0.038 &0.050\\
12I & 614 & -0.022 & 0.064 & 0.173 & 0.055 & 0.063 & 0.059&0.042\\
13C & 614 & -0.007 & 0.046 & 0.189 & 0.040 & 0.035 & 0.035&0.072\\
\hline
median(all)& 614 & -0.001& 0.036 & 0.157 & 0.037 & 0.033 & 0.032&0.062\\
median(5)  & 614 & -0.005& 0.041 & 0.128 & 0.033 & 0.028 & 0.027&0.073\\
\tableline
\end{tabular}
\tablecomments{
See comments for Table \ref{table2}.
}
\label{table3}
\end{table*}

Presenting photometric redshift accuracy as the full scatter, $\sigma_F$, gives a non-optimal representation of the scatter since a few objects (i.e., outliers) can drive the scatter to large values. Therefore, the scatter in photometric vs. spectroscopic redshifts is often expressed as the rms after excluding outliers. With this approach there are two quantities that together determine how well a code works, the rms after excluding outliers and the fraction of outliers.

The tables show that most codes produce results that broadly agree. The scatter after excluding outliers is typically in the range $\sigma_O\sim 0.04-0.07$~and the outlier fraction (OLF) is within the range 0.04-0.07 for a majority of the codes. Codes with low $\sigma_O$~tend to have a low OLF. Comparing the scatter $\sigma_O$~using the fixed outlier definition with the scatter $\sigma_{dyn}$ (which uses the dynamic outlier definition), shows a very similar rank between methods; codes with small  $\sigma_O$~have small $\sigma_{dyn}$~and codes with high $\sigma_O$~also have high $\sigma_{dyn}$. The outlier fraction is naturally less correlated between the methods. By definition, codes with $\sigma_{dyn}>0.05$~will get a lower dynamic outlier fraction, OLF$_{dyn}$, compared to the fixed outlier fraction OLF and vice versa. Due to the similarity in both the size and rank between the results of the two definitions, we will quote results using $\sigma_O$ and OLF as default, but will also include results from the dynamic definition. The fixed definition allows for comparisons between results and the literature.

In overall performance, there are five codes that have a combination of both low scatter and outlier fraction for both catalogs, i.e., codes 3B, 6E, 7C, 11H, and 13C. Inspecting Table \ref{table1}, reveals that these five results represent four different photometric redshift codes and four different sets of template SEDs. The result that no particular code gives a significantly better results than others is not surprising since most codes, including the four resulting in the lowest scatter, are based on the same technique, the $\chi^2$~template fitting. Maybe a bit more surprising is that four (or almost five) different SED sets are represented, indicating that there is not a preferred set. We note that all five codes use the training sample of galaxies to derive zero-point shifts and/or corrections to the template SED set used. Furthermore, all five include templates with emission lines and perform additional smoothing of the given flux errors. This suggests that having a code with these options is important for the quality of derived photometric redshifts. Finally, all these codes use template SEDs that include emission lines features, suggesting their importance when deriving photometric redshifts.

At the other end of the spectrum, there are a few codes with an elevated fractions of outliers compared to the other codes. For the $H$-band selected sample (Table \ref{table2}), codes 2A, 5D, and 8F have a slightly higher outlier fraction. For code 5D, this should mainly be due to the lack of templates matching low luminosity galaxies. This drives the outlier fraction to high values. Furthermore, it also prevents the code from converging for many fits, resulting in derived photo-z for only a fraction of the objects in the catalog (about 30 \% lack a calculated photo-z). For the other two codes, the higher outlier fractions could be due to a combination of not adding smoothing errors, lack of training (i.e., deriving zero-point offsets), or a limited parameter space for constructing the template SED sets. There are two codes, 5D and 8F with a resulting scatter  $\sigma_O>0.05$ in the H-band selected catalog and  $\sigma_O>0.06$ in the z-band selected catalog. For code 5D, the mass limit on the template SEDs used in this investigation should be the driving factor behind the high scatter. We also note that neither code 5D or code 8F use the spectroscopic training sample to optimize results.

For the three participants that use the EAZY code, the spread in results is comparable to the other codes that also use traditional $\chi^2$~fitting. The scatter should be due to the differences in templates, training of the code, and priors used between the participants running EAZY, which are the main parameters that vary between any $\chi^2$~fitting code, as discussed in Section 3.

From the descriptions of the different codes in Section 3, it is clear that there are many different approaches for treating data points with negative fluxes. For the spectroscopic training sample, the galaxies are relatively bright and there are few data points that are ``non-detections'' with negative fluxes ($\sim$1\%). Therefore, the different treatment of negative fluxes will likely not introduce any extra scatter or biases between codes. At finter limits though, this may lead to systematic differences between the output of the codes.

In Table \ref{table2} and \ref{table3}, we also list the scatter between the photometric redshifts and spectroscopic redshifts where we adopt the median photometric redshift from all codes and the median from the five codes with the lowest scatter.
It is very interesting that taking the median in this way produces a lower scatter and outlier fraction than any of the individual codes. This important result is discussed in Section 5, where we investigate different approches of combining results to improve the photometric redshift accuracy.

\begin{figure*}
\epsscale{2.0}
\plotone{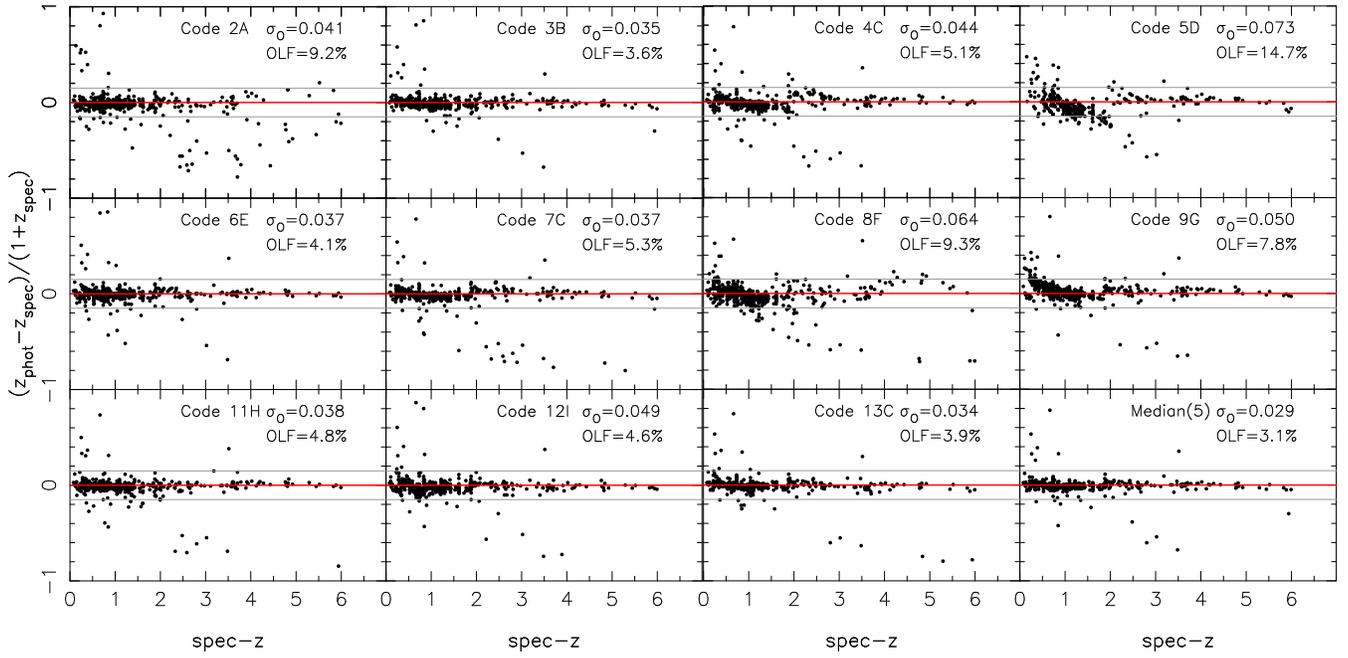}
\figcaption[f1.eps]{Comparison between photometric and spectroscopic redshifts for a sample of 589 WFC3 $H$-band selected galaxies with highest quality spectra. Figure shows codes as listed in Table \ref{table1}. Bottom right panel shows the result after taking the median of the five codes with the lowest scatter.
\label{fig2}}
\end{figure*}

To illustrate how well the individual codes recover the redshifts of the spectroscopic sample, we plot in Figure \ref{fig2}~$(z_{phot}-z_{spec})/(1+z_{spec})$~vs. $z_{spec}$~for each code for the $H$-band selected catalog. Also plotted in the right bottom panel in the figure is the scatter after calculating the median of the five codes with the lowest scatter. 

\begin{figure*}
\epsscale{1.1}
\plotone{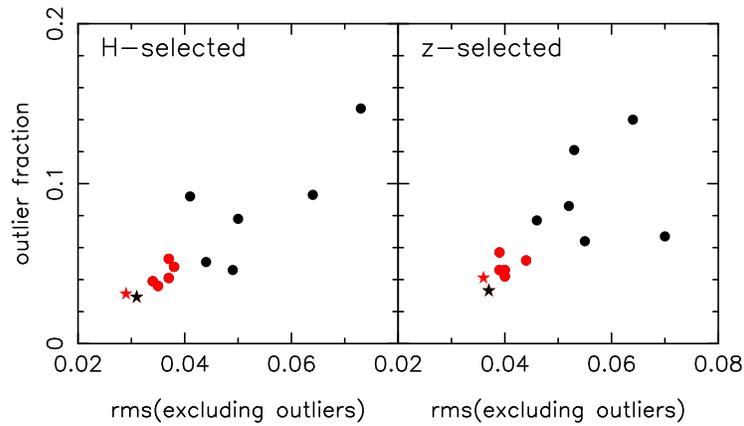}
\figcaption[f3.eps]{The rms after excluding outliers ($\sigma_O$) and outlier fractions for the different codes. The five codes with the lowest combination of scatter and outlier fractions are plotted in red. Black star symbols show the median of all codes, while the red stars show the median of the five codes with the smallest scatter. 
\label{fig3}}
\end{figure*}

To compare the results for all codes, we plot in Figure \ref{fig3}, the rms, $\sigma_O$, together with the outlier fraction for all codes for both catalogs. In red, we highlight the five codes that produce the lowest scatter and outlier fraction (i.e., are located closest to the origin). Besides the individual results, we also plot the median photometric redshift of all codes (black star symbol) and the median of the five codes with the smallest scatter  (red star symbol). This illustrates that taking the median decreases both the scatter and fraction of outliers.

\begin{figure*}
\epsscale{1.1}
\plotone{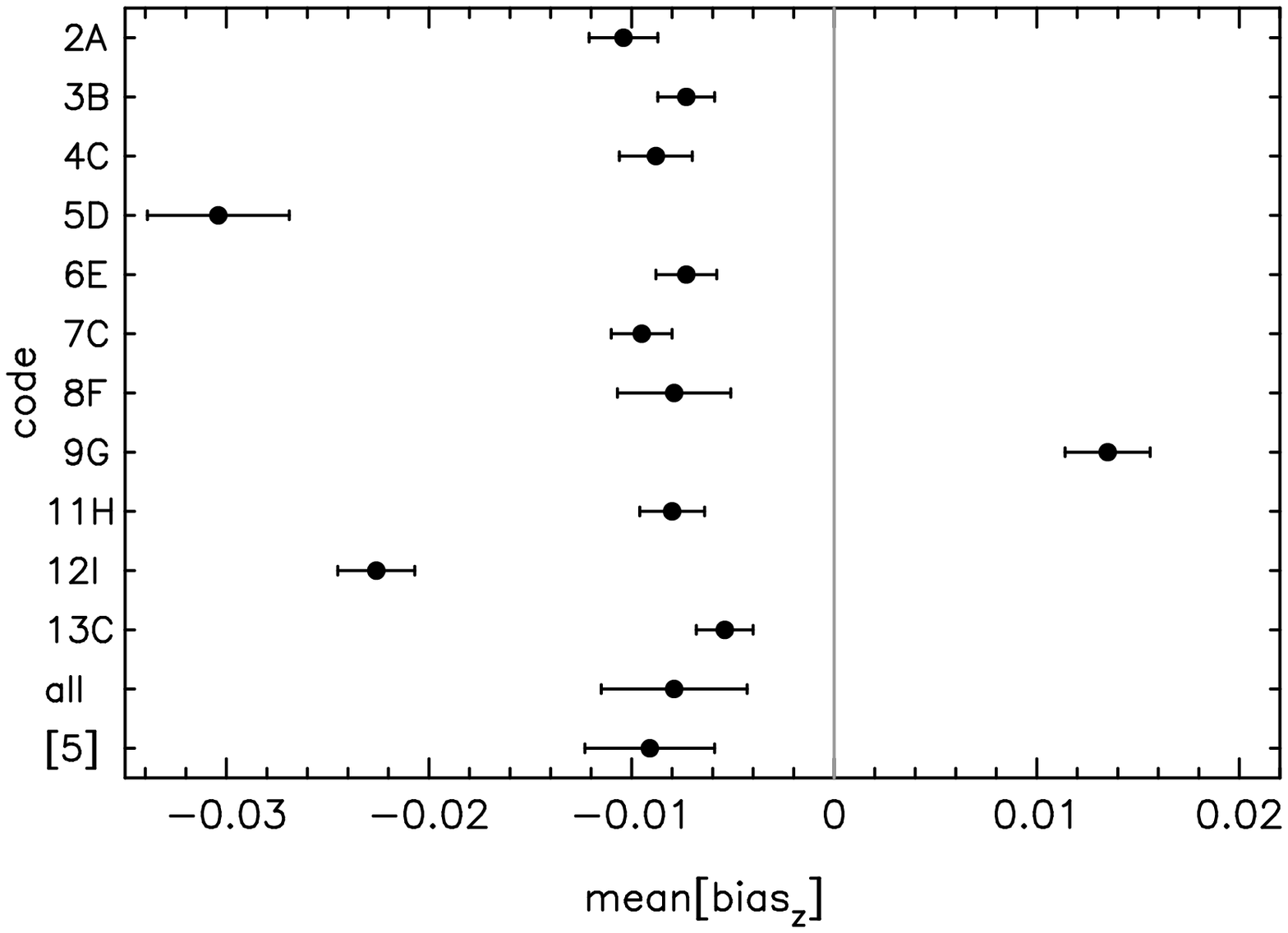}
\figcaption[f4.eps]{The mean $bias_z$~in the photometric  redshift determinations for the $H$-selected catalog. Results are shown for all individual codes, as well as the median of all codes and the median of the five codes with the smallest scatter. Error bars represent the error in the mean.
\label{fig4}}
\end{figure*}

In Figure \ref{fig4}, we plot the mean bias for the different codes, as well as for the median of all codes and the five selected codes. We find most codes produce photometric redshifts that are slightly shifted by mean[$\Delta z/(1+z_{spec})$]$\sim$0.01 in a sense that the photometric redshifts predict higher redshifts compared to the spectroscopic sample. Calculating the error in the mean as $\sigma_{bias_z}/\sqrt{N}$, where $N$~is the number of data points, we find typical errors in the mean of $\sim$0.002. Therefore, all codes have biases inconsistent with zero at a $\gsim$3$\sigma$~level. However, the biases are smaller than the scatter ($\sigma_O$) and will not dominate the overall uncertainties in the photometric redshifts. 

\subsection{Photometric redshift accuracy as a function of selection band}
Including NIR data when deriving photometric redshifts is important for photometric redshift accuracy and limiting outliers (e.g., Hogg et al. 1998; Rudnick et al. 2001; Dahlen et al. 2008, 2010). Therefore, having a catalog selected in the NIR should in principle be better than an optically selected catalog since the former assures the availability of NIR data. Of course, having an optically selected catalog that requires NIR coverage should be as close to equivalent to an NIR selected. If we compare the results from the WFC3 $H$-band selected catalog (Table \ref{table2}) with the ACS $z$-band selected catalog  (Table \ref{table3}), we find that the scatter is similar for each code. This is not unexpected since most of the photometry in the two cases are based on the same images, only the NIR bands differ. In more detail, the scatter for 9 of the 11 codes and the outlier fraction for 7 of the 11 codes are lower in the $H$-band selected catalog compared to the $z$-band selected. This slight improvement is consistent with the expected better performance for a NIR selected catalog combined with the extra depth and number of bands when replacing ISAAC $J$~and $H$~by WFC3 F098M/F105W, F125W and F160W. 

The bias$_z$~shows similar trends in the two catalogs, with deviations that are statistically inconsistent with being zero, but the absolute values are small compared to the scatter, $\sigma_O$.

Since the CANDELS survey is foremost an infrared survey for which planned catalogs are to be selected in the WFC3 infrared bands, we will concentrate our investigation on the $H$-band selected galaxy sample.

\subsection{Photometric redshift accuracy as a function of magnitude}
It is important to note that the photometric redshift accuracy reported for any survey may not be representative of the actual sample of galaxies for which photometric redshifts are derived. The reason being that the scatter is calculated using a subsample of galaxies with spectroscopic redshifts that in most cases are significantly brighter, and in many cases at lower redshift, compared to the full galaxy sample. Since fainter galaxies have larger photometric errors and may be detected in fewer bands, we expect that the errors on the photometric redshifts increase for these objects (e.g., Hildebrandt et al. 2008). As an example of the magnitude and redshift dependence on the photometric redshifts, Ilbert et al. (2009) report for the COSMOS survey $\sigma_{NMAD}$=0.007 and OLF=0.7\%~for a sample of galaxies at redshift $z<1.5$~brighter than $i^+_{\rm AB}=$22.5. At fainter magnitudes and higher redshift, they report $\sigma_{NMAD}$=0.054 and OLF=20\%~for galaxies with  redshift $1.5<z<3$~brighter than $i^+_{\rm AB}\sim$25, illustrating the significance of this effect.

To quantify the magnitude dependence of the photometric redshifts, we divide the spectroscopic sample from the $H$-band selected catalog into two magnitude bins with equal number of objects, one brighter and one fainter than $m(H)$=22.3. We find that the scatter in the median photometric redshift increases from $\sigma_O$=0.027 to $\sigma_O$=0.034 and the outlier fraction decreases from 3.1\% to 2.7\% when going from the bright to the faint subsample. The difference is small, reflecting the relative brightness of both subsamples. As a comparison, we find the that faint spectroscopic subsample has a median $m(H)$=23.2, significantly brighter than the median magnitude of the full sample, which is $m(H)$=25.7.

\begin{figure*}
\epsscale{1.4}
\plotone{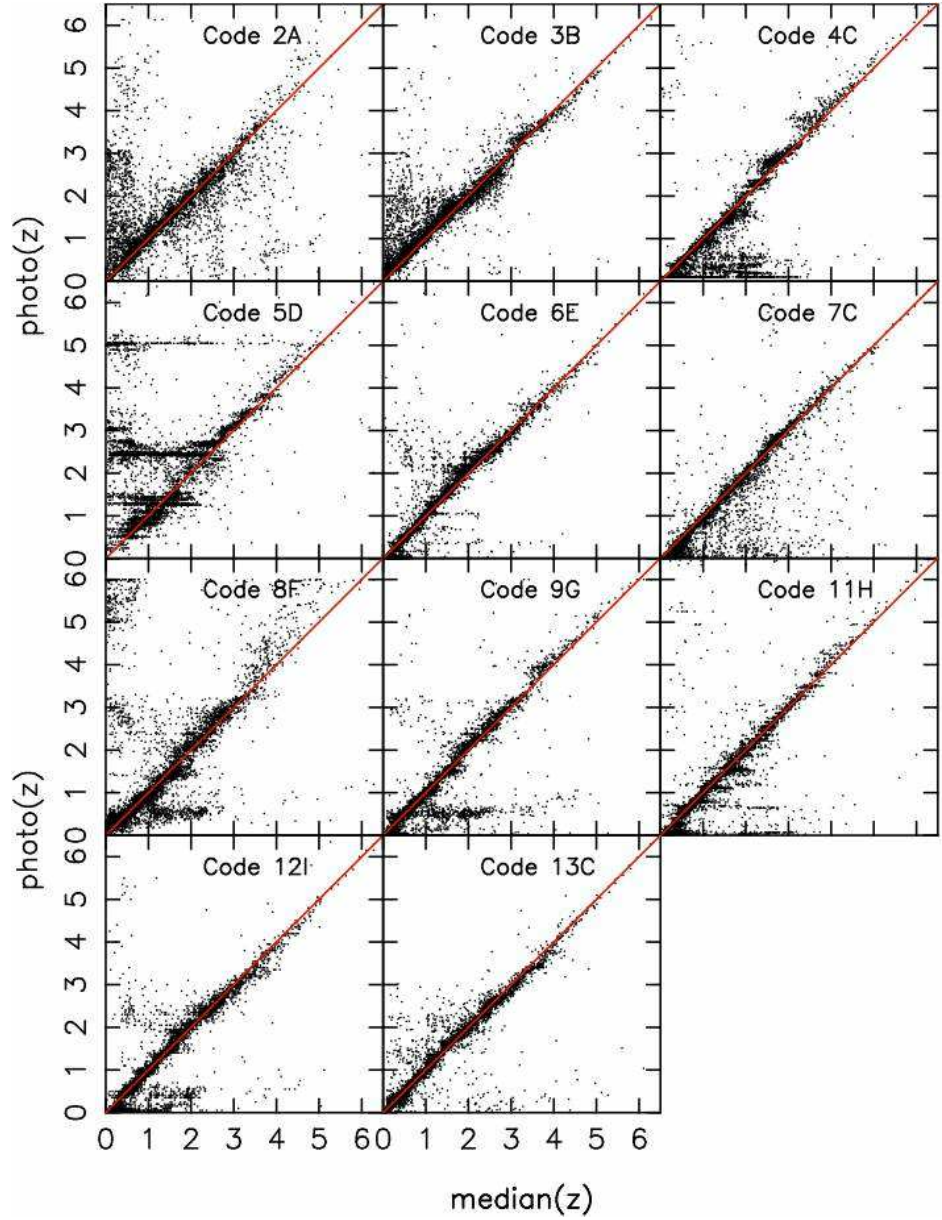}
\figcaption[f5.eps]{Scatter between individual codes and the median of all eleven codes using the $H$-selected catalog with signal-to-noise $>$10. 
\label{fig5}}
\end{figure*}

To visualize the behavior of photometric redshifts down to faint magnitudes, we plot in Figure \ref{fig5}  the scatter between the eleven individual codes and the median of all codes. Each panel shows about $\sim$6000 objects with signal-to-noise $>$10. We do not know how well the median represents the true redshifts at these magnitudes, but the plot illustrates that there are some substantial biases in a number of codes. For example, codes 2A, 5D, and 8F have a fairly prominent population at higher redshift compared to the median. Potentially due to the aliasing between the Lyman and the 4000\AA~breaks these codes more often chose the higher redshift solution compared to the median. Again, we note that the median we compare to is not necessarily the most correct solution.
%%%

\begin{figure*}
\epsscale{1.4}
\plotone{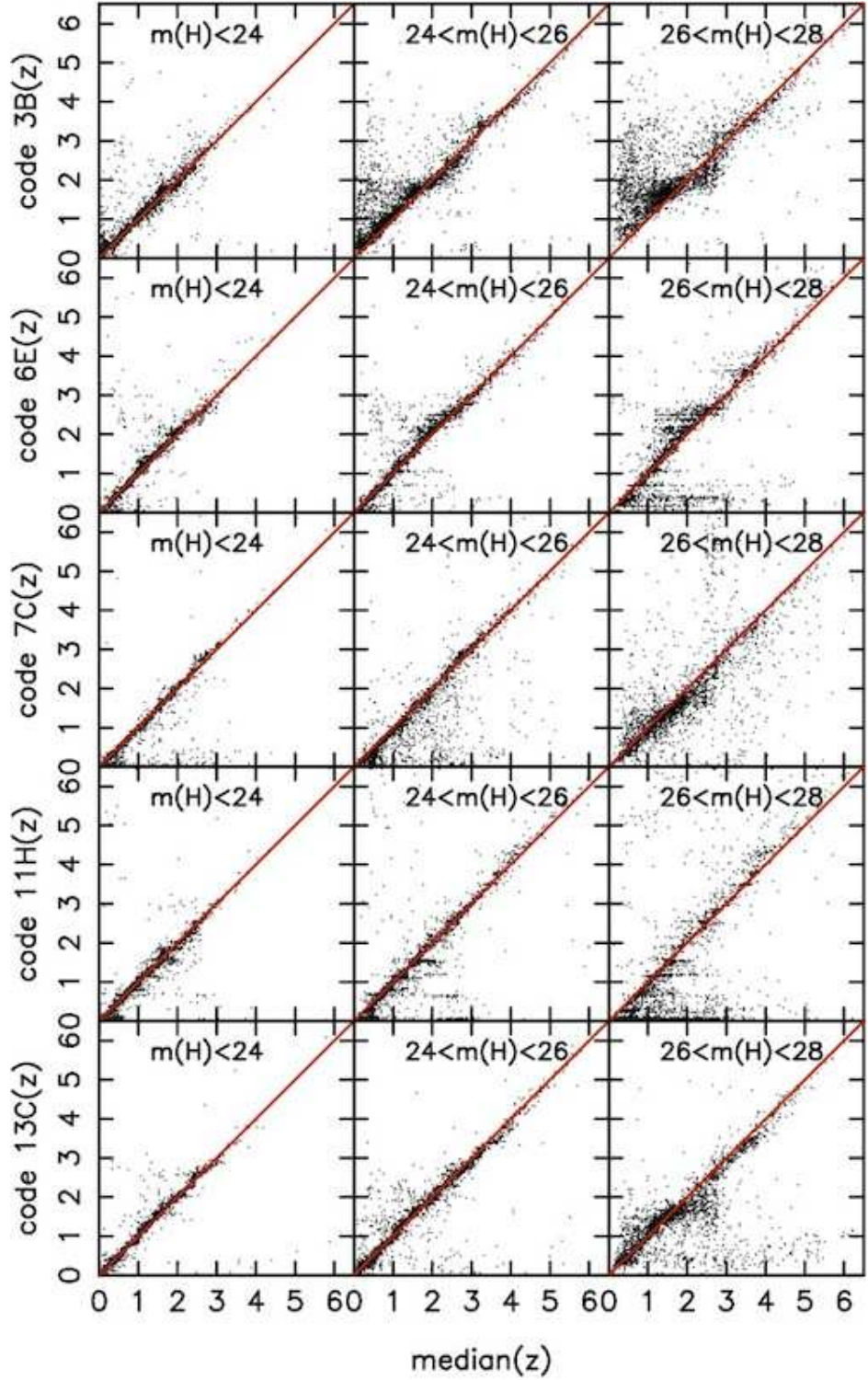}
\figcaption[f6.eps]{Scatter between the five individual codes with the lowest scatter (codes 3B, 6E, 7C, 11H, and 13C) and the median of all 11 codes. Each column show a different magnitude selection $m(H)<24$, $24<m(H)<26$, and $26<m(H)<28$. The same number of objects are shown in each panel.
\label{fig6}}
\end{figure*}

To check the magnitude dependence for the full galaxy sample in some more detail, we plot in Figure \ref{fig6}, the comparison between the five codes with the lowest scatter (3B, 6E, 7C, 11H, and 13C) and the median of all codes in three magnitude bins, $m(H)<$24, 24$<m(H)<$26, and 26$<m(H)<$28. It is clear from the figure that the scatter increases at fainter magnitudes (note that we plot the same number of objects, $\sim$3000, in each panel). To quantify the magnitude dependence, we calculate the mean scatter between the individual codes and the median in the three magnitude bins and find $\sigma_O$=0.040, 0.048, and 0.055, respectively. For the fraction of outliers, we find for the three bins OLF=8\%, 16\% and 28\%, respectively. This increase in scatter, and particularly in the fraction of outliers, further illustrates that the dispersion in the photometric redshifts calculated by different codes becomes significant at faint magnitudes, even though a good agreement is noticeable at brighter magnitudes. Interestingly though, there is a fairly good agreement between code at all magnitudes at redshifts $z\gsim$3-4. This should be due to the strong Lyman-break feature at these redshifts that helps determine the photometric redshift. We select these five particular codes because at bright magnitudes (i.e., typical of the spectroscopic samples) they produce very similar photometric redshifts. This allows us to investigate how results diverge between codes due to mainly the signal-to-noise. We made similar tests using different codes and find results that are consistent.

\subsubsection{Simulating a faint spectroscopic redshift sample}
\begin{figure*}
\epsscale{1.1}
\plotone{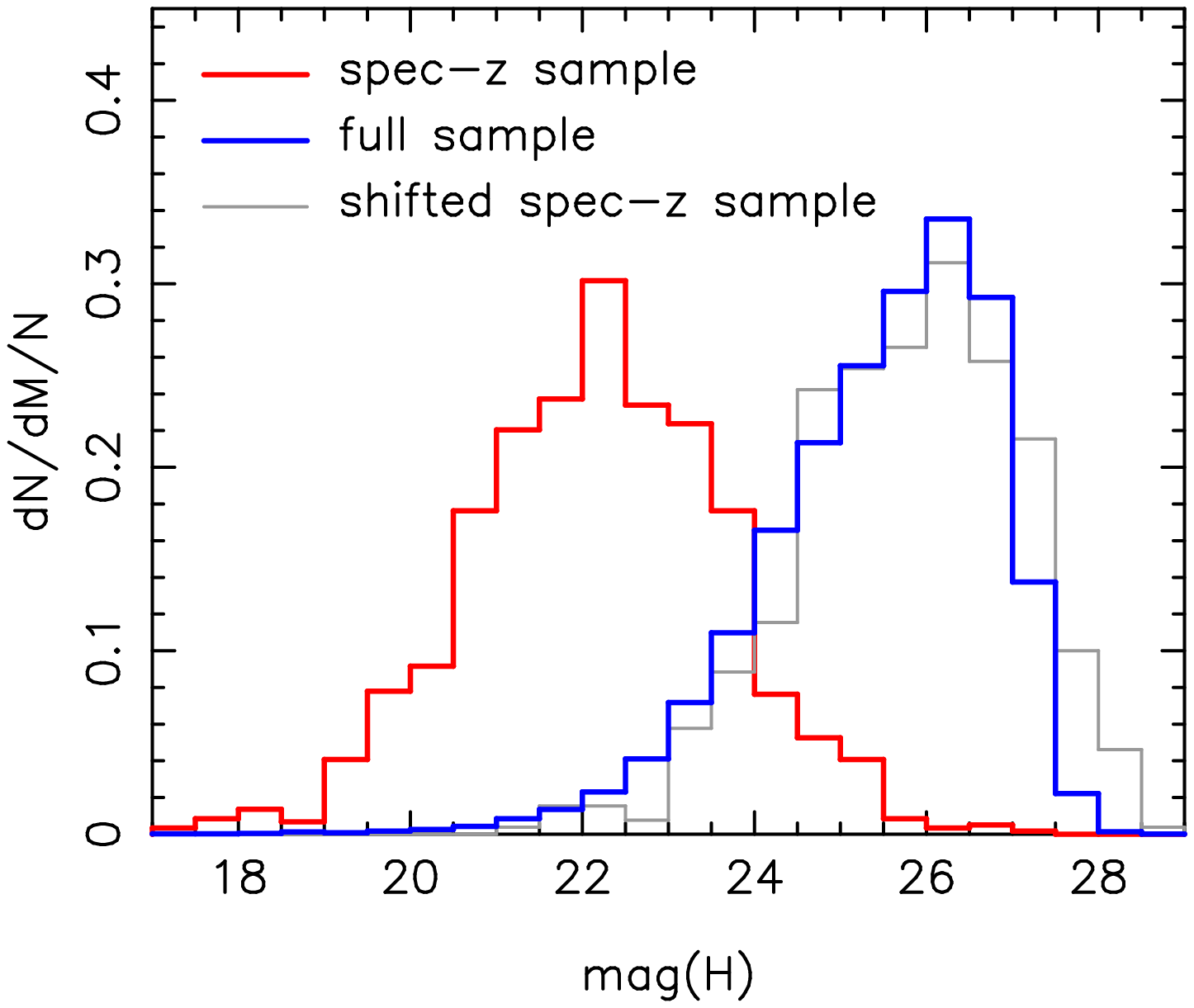}
\figcaption[f7.eps]{Magnitude distribution of the spectroscopic sub-sample of GOODS-S is shown in red while the full sample is shown in blue. Gray line shows the degraded spectroscopic sample where the flux of each object has been shifted by $\Delta m$=3.6 mag to match the full sample. The distributions are normalized to the total number of objects in each sample.
\label{fig7}}
\end{figure*}

To quantify the difference between the brightness distribution of the sub-sample with spectroscopic redshifts, compared to a full galaxy sample, we plot in Figure \ref{fig7} the normalized distributions of the available spectroscopic sample together with the full sample of galaxies for the GOODS-S $H$-band selected catalog. For the full sample, we restrict the selection to galaxies with S/N$>$5 in the $H$-band that are detected in at least six photometric bands. The red line in the figure shows the distribution of the spectroscopic sample while the blue line shows the full sample. Obviously, the spectroscopic sample is significantly brighter than the bulk of the full sample of galaxies. When using the S/N$>$5 limit in the $H$-band, we find that the full sample is on average 3.6 mag fainter than the spectroscopic sub sample.

To better quantify how the brightness of the spectroscopic sample affects the photometric redshift accuracy, we artificially make the spectroscopic sample fainter to resemble the flux distribution expected for a deeper spectroscopic sample. First, we make a catalog consisting of the $\sim$ 1000 objects with highest quality spectroscopic redshifts from the $H$-band selected catalog. The catalog initially has a magnitude distribution according to the red line in Figure \ref{fig7}. We thereafter make all fluxes fainter by $\Delta m$=3.6 mag, which is the average difference between the spectroscopic sample and the full sample in Figure \ref{fig7}. To each new flux value we assign a photometric error drawn from the original catalog at a flux level matching the new assigned flux. We finally perturb the flux values using the assigned errors, assuming that they are Gaussian and represent 1$\sigma$. The new magnitude distribution of the shifted spectroscopic sample is shown by the gray line in Figure \ref{fig7}. This distribution is consistent with the distribution of the full photometric sample. To further quantify the flux dependence of the photometric redshifts, we have also made catalogs where we shift the spectroscopic sample by $\Delta m$=1, 2, 3, and 4 mag, respectively

\begin{figure*}
\epsscale{1.1}
\plotone{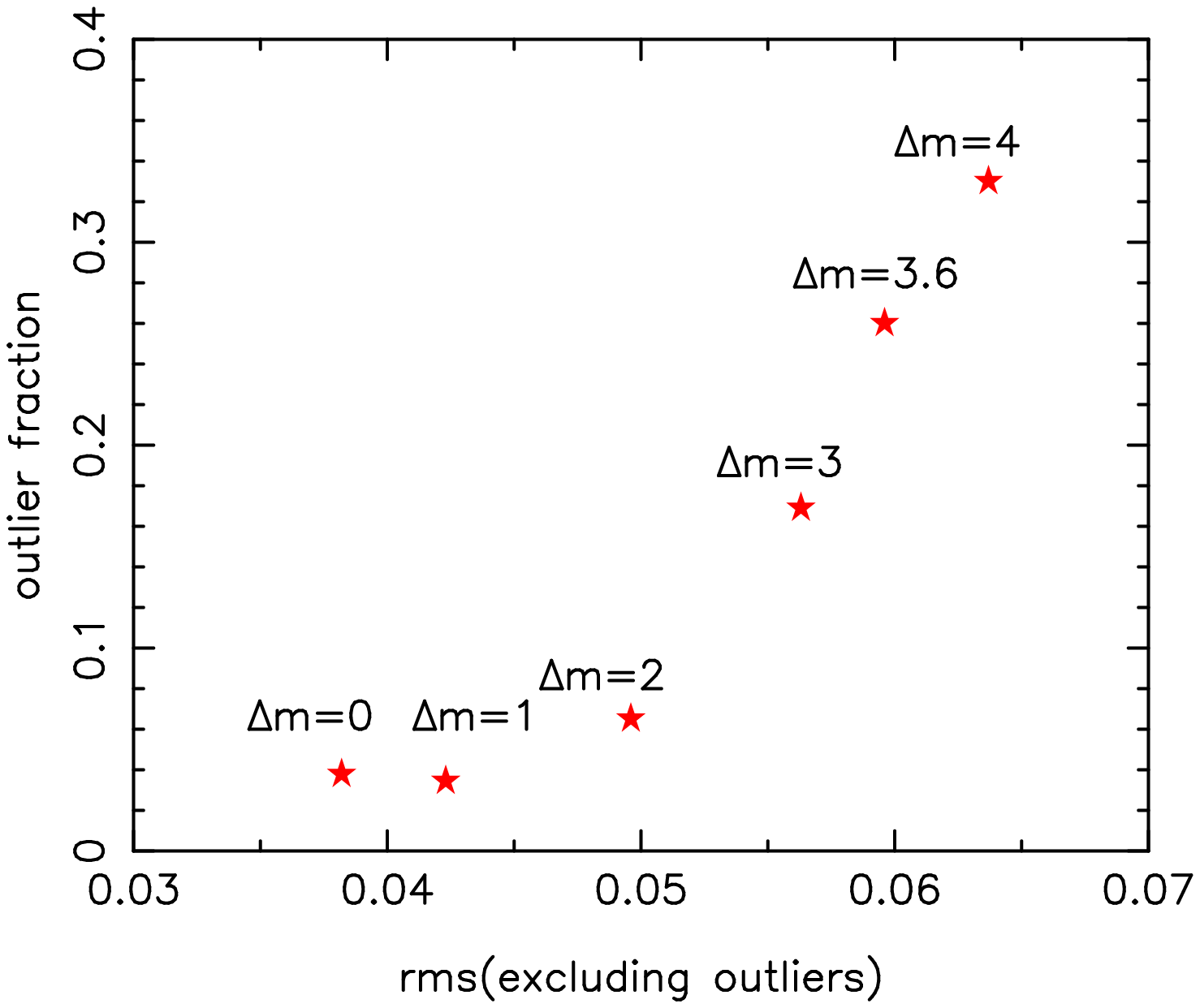}
\figcaption[f8.eps]{Photometric redshift scatter ($\sigma_O$) and outlier fraction when comparing to nominal spectroscopic redshift sample ($\Delta$m=0), as well as samples where the photometry as been shifted to fainter flux levels by $\Delta m$=1, 2, 3, 3.6, and 4 mag, respectively. Results are shown for one participating code (Code 3B).  
\label{fig8}}
\end{figure*}

To show the flux dependence of the photometric redshift accuracy, we plot in Figure \ref{fig8} the scatter and outlier fraction for the nominal case and for the five catalogs with perturbed photometry. We illustrate results from one specific code (Code 3B), but we expect a similar behavior for all codes. It is clear that both the scatter and outlier fractions increase as the spectroscopic sample is shifted to fainter flux levels. Particularly, there is a significant increase in outlier fraction at faint magnitudes $\Delta m\gsim 2$. This could be related to the increased risk of misidentifying the Lyman and 4000\AA~breaks at fainter magnitudes where photometric error are larger. 

\begin{figure*}
\epsscale{1.1}
\plotone{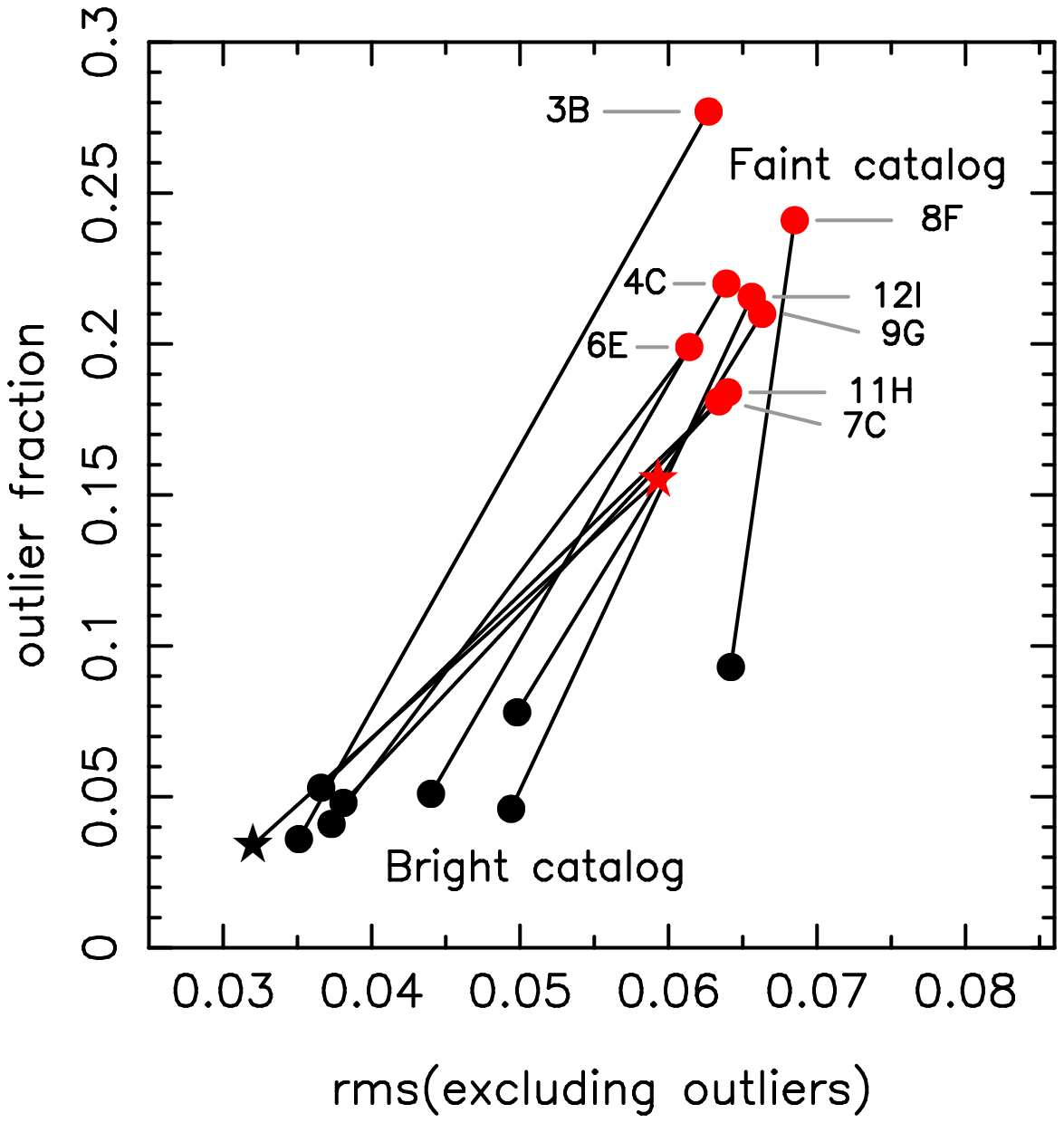}
\figcaption[f9.eps]{Photometric redshift scatter ($\sigma_O$) and outlier fraction for individual codes. Black dots show results from the original $H$-band selected catalog, while the red dots show the results after fluxes are shifted to fainter limits by $\Delta m$=3.6. Lines connect the results from the separate codes. Star symbols show the results when using the median of the photometric redshifts of the eight codes participating in this test.
\label{fig9}}
\end{figure*}

In a second test using the shifted photometry of the spectroscopic sample, we compare the results from multiple codes run on the same catalog. Here we use the $\Delta$m=3.6 catalog, since this illustrates the difference in photometry between the spectroscopic catalog and the full $H$-band selected catalog used in this investigation. Eight codes participated in this test (codes 3B, 4C, 6E, 7C, 8F, 9G, 11H, and 12I). Results are shown in Figure \ref{fig9}. Black dots to the lower left show the photometric redshift scatter and outlier fraction for the original case, while red dots in the upper left show the results after shifting the catalog to fainter fluxes and increased errors. Star symbols represent the results when using the median of all codes. Obviously, both the scatter and outlier fraction increase significantly for all codes when the photometric errors increase. For the median case, the scatter approximately doubles from $\sigma_O$=0.03 to $\sigma_O$=0.06, while the outlier fraction increases from 4\% to 15\%. At the same time, we note that in the case with shifted photometry, the median produces better results than any of the individual codes.

\begin{figure*}
\epsscale{1.3}
\plotone{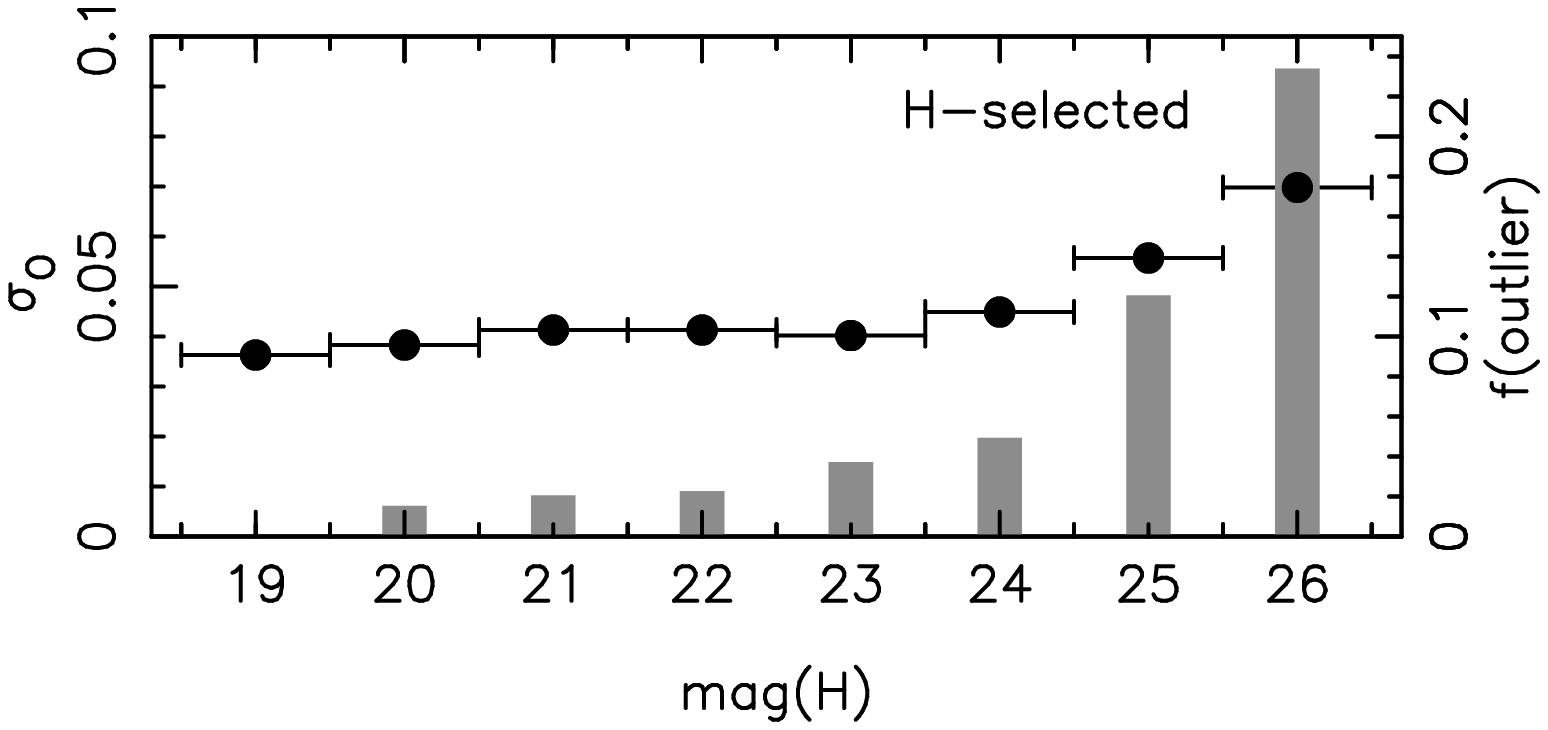}
\figcaption[f10.eps]{The magnitude dependence of the photometric redshift scatter and outlier fraction using photometric redshifts derived from a mock catalog based on the spectroscopic redshift sample shifted to fainter magnitudes. Black dots show the scatter $\sigma_O$~(scaling on left-hand $y$-axis, error bars show bin size). Histograms show the fraction of outliers (scaling on right-hand $y$-axis).
\label{fig10}}
\end{figure*}

As a final test, we use data from the simulated catalogs that were made artificially fainter to investigate the reliability of the photometric redshifts as a function of magnitude, using one of the codes (Code 3B), as a representative case. In Figure \ref{fig10} we show the scatter ($\sigma_O$) and outlier fraction in magnitude bins with $\Delta m=1$~over the range $19<m(H)<26$. The Figure indicates that both the scatter and outlier fractions are reasonably well behaved and degrade slowly out to magnitudes $m(H)\sim 24$, whereafter both quantities increase more rapidly at $m(H)\gsim 25$ . 

\subsection{Photometric redshift accuracy as a function of redshift}
\begin{figure*}
\epsscale{1.3}
\plotone{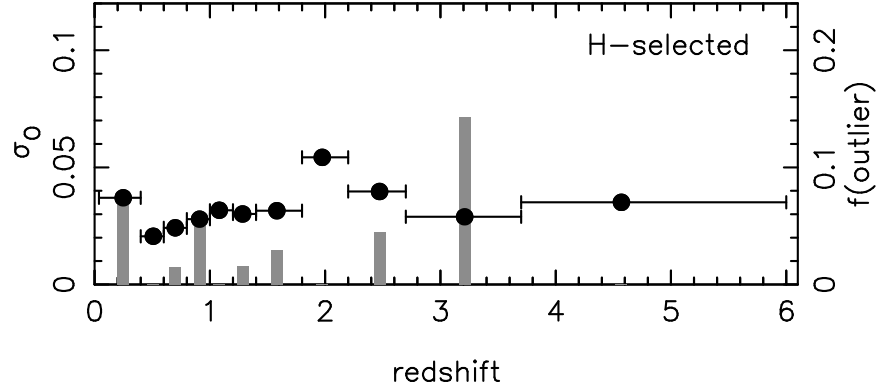}
\figcaption[f11.eps]{Redshift dependence of the photometric redshift scatter and outlier fraction when comparing the median photometric redshift with the spectroscopic redshift sample. Black dots show the scatter $\sigma_O$~ (scaling on left-hand $y$-axis). Histograms show the fraction of outliers (scaling on right-hand $y$-axis).
\label{fig11}}
\end{figure*}

To test the redshift dependence of the photometric redshifts, we first divide the spectroscopic control sample in the $H$-band into two bins with equal number of objects. The redshift dividing the bins is $z_{spec}$=0.95 and the median redshift for the two bins are $z_{spec}$=0.7 and $z_{spec}$=1.4, respectively. We find that the scatter in the median photometric redshift increases from $\sigma_O$=0.027 to $\sigma_O$=0.034, while the outlier fraction decreases from 3.4\% to 2.4\% when going from the low redshift to the high redshift subsamples. This indicates that there is no strong redshift trend in the photometric redshift accuracy. To make a more detailed investigation, we divide the spectroscopic sample into eleven redshift bins and calculate the scatter and outlier fraction in each bin separately. Figure \ref{fig11} shows the result for the $H$-band selected catalog when comparing the median photometric redshifts to the spectroscopic redshifts. The scatter, $\sigma_O$, lies at a fairly constant level with redshift, indicating that the redshift-normalized scatter  gives a fairly robust indicator of the photometric redshift accuracy almost independent of redshift. The only point that lies significantly above the trend is the $z\sim$2 point. This could be due to the lack of strong spectral features at this redshift. This is also the redshift range where we expect the spectroscopic redshifts to be most uncertain and we cannot rule out some errors in the spectroscopic sample even though we limit our selection to the highest quality spectra. At higher redshifts, the Lyman break moves into the $U$-band, providing an important signal for the photometric redshift determination (e.g., Rafelski et al. 2009). We also note that the VIMOS $U$-band used is redder than the typical $U$-band and therefore starts to probe the break at slightly higher redshifts. Possibly contributing to the relatively high outlier fractions in the $z\sim$2.5 and $z\sim$3.2 data points. However, the tests do not account for high-z galaxies with significantly different SEDs than the moderate-z spec sample. If such a population is common at high redshift and is unrepresented in the template SED libraries, it could affect the accuracy of the photometric redshifts. It is, however, reassuring that for the spectroscopic sample at $z>3\sim 4$, the photometric redshifts agree well with the redshift from the spectra (e.g., Figure \ref{fig2}). Contributing to the accuracy of the $z>3$ photometric redshifts is the break due to absorption by intergalactic HI clouds (Madau 1995), which affects the observed signal for all galaxy SED types. In fact, in Figure \ref{fig11}, there are no outliers in the highest redshift bin ($z>3.7$), indicating that the Lyman break helps to provide robust photometric redshift determinations at these redshifts.

\subsection{Photometric redshift accuracy as a function of galaxy spectral type}
The most important spectral features for determining photometric redshifts are the Lyman break at $\sim$ 1215\AA~and the 4000\AA~break (we let the 4000\AA~break denote the overall spectral feature caused by the Balmer break at 3646\AA~and the accumulation of absorption lines of mainly ionized metals around $\sim$4000\AA). It is also expected that the size of the break should be important for the accuracy of the photometric redshifts. For example, an old red galaxy with a pronounced 4000\AA~break should result in more accurate photometric redshift compared to a younger blue galaxy with a more featureless SED. These effects should be most important at lower redshifts ($z\lsim 2-3$), where the redshifted Lyman break has not yet entered the observed $U$-band. At higher redshifts where the break at rest frame wavelengths short of $\sim$1215 \AA~ (Madau 1995) moves into the observed bands, even intrinsically featureless blue galaxies will show a break feature that helps determine the photometric redshift. Galaxies with blue, relatively featureless SED at redshift $z\lsim 2-3$~therefore have the highest risk of being assigned incorrect redshifts.

\begin{figure*}
\epsscale{1.3}
\plotone{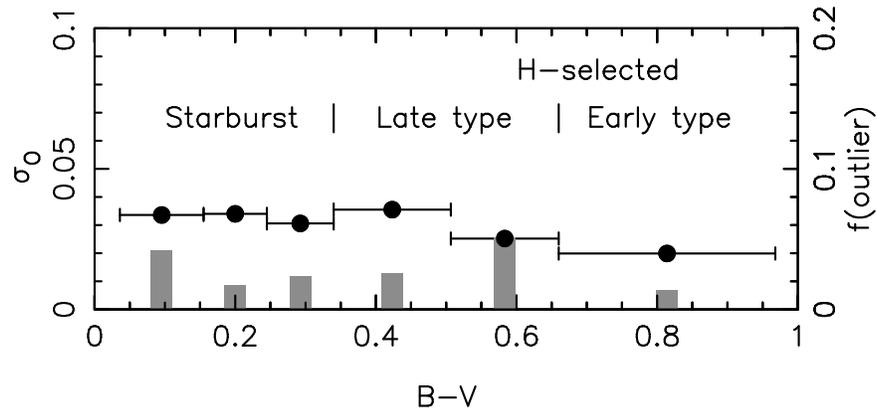}
\figcaption[f12.eps]{The photometric redshift scatter and outlier fraction when comparing the median photometric redshift with the spectroscopic redshift sample as a function of galaxy color. Black dots show the scatter $\sigma_O$~ (scaling on left-hand $y$-axis). Histograms show the fraction of outliers (scaling on right-hand $y$-axis).
\label{fig12}}
\end{figure*}

To investigate the photometric redshift accuracy as a function of galaxy spectral type, we divide our spectroscopic sample into early-types, late-types, and starbursts based on the rest frame $B-V$~color of the galaxy. The colors are calculated using the observed bands that most closely covers the redshifted rest frame $B-V$~together with K-corrections based on the best-fitting template SED following the method in Dahlen et al. (2005). We use a division where galaxies with $B-V <0.34$~are assigned as starbursts and galaxies with $B-V >0.66$~are assigned as early-type galaxies. Galaxies with intermediate colors are assigned as late-type galaxies. This is a single rest frame color definition, we can not rule out that dusty later type galaxies may fall into the early-type category. In Figure \ref{fig12}, we plot the scatter and outlier fraction for the control sample in six color bins using the median photometric redshift when comparing to the spectroscopic redshift.

From the figure we note that the scatter is not strongly dependent on galaxy color. There is an indication that the early-types have a smaller scatter ($\sigma_O \sim$0.02) compared to the remaining types ($\sigma_O \sim$0.03). However, within the starburst and late-type bins, there is no clear color dependence. If we exclude galaxies at $z>2$~(where the Lyman break may be useful for determining photometric  redshifts) there is no significant change in the results. We therefore conclude that there is no strong color dependence in the photometric redshifts, except that we may expect more secure redshifts for early-type galaxies.

\subsection{Applying zero-point shifts and smoothing errors}
The five codes resulting in the lowest scatter and outlier fraction use the spectroscopic training sample to derive shifts to either the photometry or template SEDs and add extra smoothing errors. The better behavior when applying zero-points shifts could be due to a number of factors. There could be actual errors in the given zero-points used to calculate the photometry, there could also be a mismatch between the template SEDs and the true SEDs of the observed objects. Furthermore, insufficient knowledge of the system throughput given by the filter transmission curves may cause offsets between observed and predicted fluxes. Finally, when photometry from different images are merged to a common catalog, there could be unaccounted aperture corrections contributing to offsets between filters. By using a spectroscopic training sample with sufficiently many objects, a number of codes offer the possibility to calculate zero-point shifts which are thereafter applied to either the photometry or the templates SEDs before deriving photometric redshifts. 

Table \ref{table4} illustrates the size of the shifts derived by codes 3B, 6E, 7C, 11H, and 13C for both the $H$-band and $z$-band selected catalogs. A positive offset in the table indicates that the observed flux is brighter compared to what is expected from the template SED. For each filter, we also give the median of the available shifts together with the error in the median. There is a noticeable scatter in the size (and sometimes sign) between the corrections derived by the different codes, suggesting that the zero-point shifts depend on the code, implementation and template SED set used. However, there are some common trends among the codes. To highlight this, we have marked in bold face the cases when the mean shift of all codes deviates from zero with at least a 5$\sigma$~significance. There are significant shifts for some of the ACS filters, even though the absolute shifts are small ($\lsim$0.03 mag). More noticeable shifts are noted for some of the NIR ISAAC filters in the $z$-selected catalog, with the $J$-band shift being significant. Most measurements indicate that the IRAC fluxes predicted by template SEDs are too faint compared to the measured fluxes. To some extent, this could be due to the lack of PAH emission at long wavelengths in many template SED libraries. 

\begin{table*}
\caption{Zero-point shifts calculated for five of the participating codes.}
\centering
\begin{tabular}{lrrrrrr}
\multicolumn{7}{c}{WFC3 $H$-selected}    \\
Filter &Code 3B & Code 6E & Code 7C & Code 11H & Code 13C & Mean \\
\tableline 
\tableline
VIMOS(U)       &0.004   &-0.013  &- &-0.033  &-0.030            &-0.018$\pm$0.007\\
ACS(F435W)   &-0.004 & 0.028  &- &0.047  &0.030             &0.025$\pm$0.009\\
ACS(F606W)   & 0.031  &0.008  &- & 0.028  & 0.032             &{\bf 0.025$\pm$0.005}\\
ACS(F775W)   & 0.010  &0.018  &- &0.002  &0.037               &0.017$\pm$0.006\\
ACS(F850LP)  & 0.010  &0.025  &- &0.015  &0.040               &0.022$\pm$0.006\\
WFC3(F098M) & -0.022&0.001  &- &0.000  &0.016                &-0.001$\pm$0.007\\
WFC3(F105W) & -0.011&0.009  &- &0.000  &0.008                &0.002$\pm$0.004 \\
WFC3(F125W) & -0.062 &-0.009  &-0.100 &-0.022  &-0.011   &-0.041$\pm$0.016\\
WFC3(F160W) & -0.091 &-0.010  &0.020 &0.005  &-0.019       &-0.019$\pm$0.017\\
ISAAC(Ks)       & -0.031& -0.013&0.020  &0.025 &-0.040   &-0.008$\pm$0.012\\
IRAC(ch1)       & 0.120   &0.117  &0.050&0.106  &0.026 &0.084$\pm$0.017\\
IRAC(ch2)      & 0.114     &0.098  &- &0.073  &-0.034 &0.063$\pm$0.029\\
IRAC(ch3)     & 0.236&0.168  &- &- &- &{\bf 0.202$\pm$0.024}\\
IRAC(ch4)     & 0.455 &0.171  &- &-  &- &0.313$\pm$0.100\\
\tableline
\multicolumn{7}{c}{ACS $z$-selected}    \\
\tableline
VIMOS(U)       & 0.018&0.029 &- &-0.027 & -0.005&0.004$\pm$0.011\\
ACS(F435W)   &-0.018 &-0.053 &- &-0.023 & -0.053&-0.037$\pm$0.008\\
ACS(F606W)   &  0.046&0.004 &- &0.016 & 0.018&0.021$\pm$0.008\\
ACS(F775W)   &  0.018&0.020 &- &0.024 & 0.025&{\bf 0.022$\pm$0.001}\\
ACS(F850LP)  &  0.018&0.027 &- &0.032 & 0.013&{\bf 0.022$\pm$0.004}\\
ISAAC(J)         & -0.095 &-0.057 & -0.050 & -0.054 & -0.094&{\bf -0.070$\pm$0.009}\\
ISAAC(H)       & -0.130&-0.060 & - & -0.010 & -0.107&-0.077$\pm$0.023\\
ISAAC(Ks)       & -0.049 &-0.006 & 0.050 & 0.091 & -0.015&0.014$\pm$0.022\\
IRAC(ch1)       &  0.101&0.131 & - & 0.175 & 0.023&0.107$\pm$0.028\\
IRAC(ch2)      & 0.083&0.105 & - & 0.111 & -0.031&0.067$\pm$0.029\\
IRAC(ch3)     &  0.198 &0.160 & - & 0.148 & -&{\bf 0.169$\pm$0.012}\\
IRAC(ch4)     & 0.351 &0.179 & - & 0.240 & -&{\bf 0.257$\pm$0.041}\\
\tableline
\tableline
\end{tabular}
\tablecomments{
Col 1: Filter, Cols 2-6: zero-point shifts for codes 3B, 6E, 7C, 11H, and 13C. Col 7: Mean shift and error in the mean. Cases when the mean deviates more than 5$\sigma$~from zero are shown in bold face. A positive shift indicates that the measured flux is too bright compared to the estimated template SED flux.
}
\label{table4}
\end{table*}

\subsection{Using pair statistics to estimate photometric redshift uncertainties}
As an alternative for estimating the photometric redshift uncertainties at faint magnitudes where spectroscopic redshifts are not available, we use the method outlined in Quadri \& Williams (2010) and Huang et al. (2013). This method uses the fact that close pairs have a significant probability of being associated and that they therefore are at similar redshifts. By plotting the distribution of differences in photometric redshifts of close pairs from the photometric redshift catalog, compared to a distribution based on any random two galaxies, the close pair distribution will show excess power at small separations reflecting an elevated probability for close pairs being at similar redshift. Here two objects are considered a close pair if the separation is less than 15 arcsec. 

\begin{figure*}
\epsscale{1.1}
\plotone{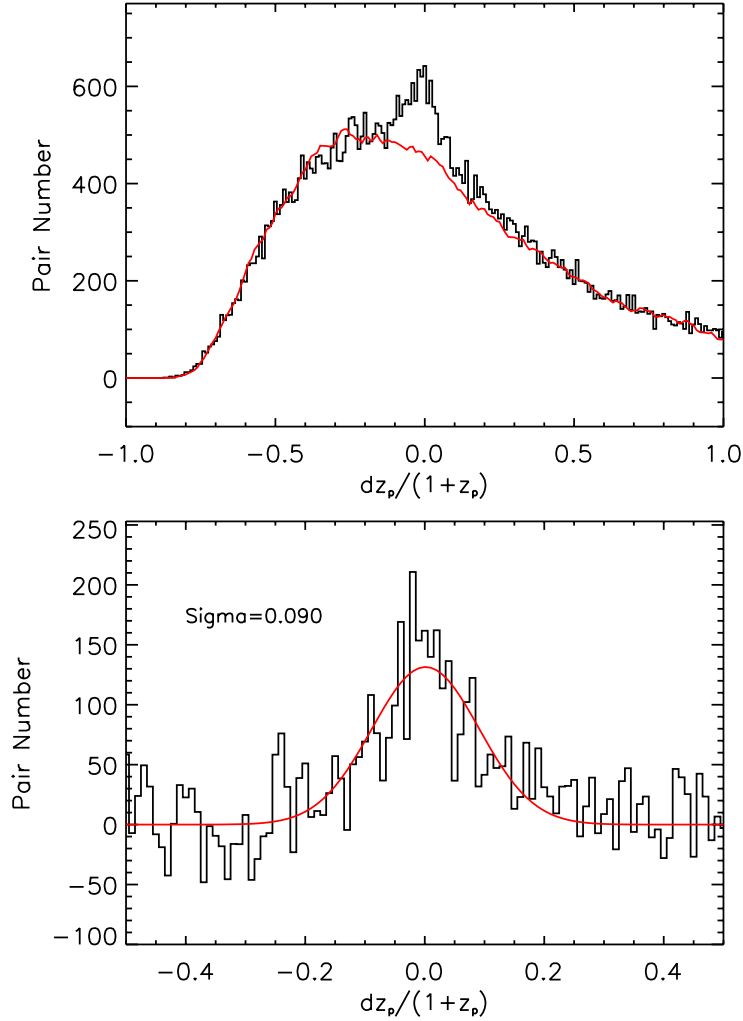}
\figcaption[gss_hpair_f1.eps]{{Top panel: distribution of difference in photometric redshifts for close pairs (black line) and random pairs (red line). Bottom panel: Overdensity of galaxy pairs with similar photometric redshifts after subtracting the random pair distribution. The red solid line is a Gaussian fit to the data. } 
\label{fig13}}
\end{figure*}

In the top panel of Figure \ref{fig13}, we show the distribution of differences in photometric redshifts for close pairs as the black line, while the red line shows the distribution for random pairs. In the bottom panel, we show the distribution of differences in photometric redshifts after subtracting out the random pair distribution. The result is shown for code 3B. Evidently, pairs with similar photometric redshifts show an excess in the distribution. Fitting a Gaussian to the excess peak in the bottom panel (red line) results in a width of $\sigma$=0.090. This width includes scatter from both galaxies in the pair for which the difference in photometric redshift is calculated. Therefore, the scatter for individual objects should be 0.090/$\sqrt{2}$=0.064. Note that only galaxies with relatively similar photometric redshifts contribute to the peak, i.e., pairs where one of the objects is an outlier will not be included. The derived width of the peak should be compared to $\sigma_O$, the scatter after excluding outliers. While the derived scatter using the close pair method is larger than the value $\sigma_O$=0.035 derived when comparing to the spectroscopic control sample, the pair method is useful to fainter limits and is not as biased towards brighter fluxes or specific galaxy types as the spectroscopic sample. For the sample shown in Figure \ref{fig13}, all galaxies with fluxes $>1\mu $Jy (corresponding to $m(H)<23.9$) are used. Going even deeper, using all galaxies with fluxes $>0.5\mu $Jy (corresponding to $m(H)<24.7$), results in a scatter  $\sigma$=0.087. These results confirm that the scatter in the photometric redshifts increases at magnitudes fainter than the spectroscopic control sample.

\subsection{Error estimates for photometric redshifts}
Most photometric redshift codes return an estimate of the uncertainty in the derived photometric redshift. This is an estimate of confidence intervals of the photometric redshifts, such as the 68.3\% and 95.4\% confidence intervals (corresponding to $\pm 1\sigma$~and  $\pm 2\sigma$~for a Gaussian distribution). There are also codes that produce full probability distributions, P(z), based on the $\chi^2$~fitting, where P(z) $\propto exp(-\chi^2)$. Ideally, these error estimates should reflect the uncertainties in the derived photometric redshifts. However, there is not necessarily a correlation between how well a photometric redshift code reproduces the spectroscopic redshifts and the accuracy of the error estimates of the photometric redshifts. Hildebrandt et al. (2008) investigated the behavior of a number of photometric redshift codes and found that the error estimates did not correlate tightly with the  photometric  redshift accuracy. As a test of how well the assigned errors reflect the actual errors, we calculate the fraction of galaxies with known spectroscopic redshifts in the control sample that falls within the 68\% and 95\% confidence intervals derived by the different codes. If quoted errors in the photometric redshifts are representative of the true redshift errors, then we expect about 68\% and 95\% of the spectroscopic redshifts fall within the two intervals, respectively. We show results in Table \ref{table5}.

\begin{table*}
\caption{Error measurement accuracies for the  $H$-band and the $z$-band selected catalogs.}
\centering
\begin{tabular}{ccccc}
Code  & \multicolumn{2}{c}{WFC3 $H$-selected}  &  \multicolumn{2}{c}{ACS $z$-selected}   \\
\tableline 
conf. int: & 68.3\% & 95.4\% & 68.3\% & 95.4\%  \\
\tableline
2A & 46.1 & ~ & 40.9 & ~\\
3B & 81.6 & 92.8 & 76.1 & 89.1\\
4C & 64.0 & 88.2 & 58.5 & 85.7\\
5D & 2.5 & 4.2 & 2.9 & 5.8\\
6E & 52.0 & 84.7 & 48.3 & 81.6\\
7C & 65.0 & 87.3 & 62.9 & 89.1\\
8F & 15.3 & 15.6 & 14.2 & 14.7\\
9G & 16.3 & 44.1 & 15.0 & 39.6\\
11H & 35.2 & 54.0$^a$ & 30.9 & 46.9$^a$\\
12I & 88.7 & 96.7 & 80.1 & 96.3\\
13C & 52.0 & 72.7 & 35.7 & 51.0\\

\tableline
\end{tabular}
\tablecomments{
$^a$ This is the result for the 90\% confidence interval. The table shows the fraction of galaxies with known spectroscopic redshifts that falls inside the 68.3\% and 95.4\% confidence intervals calculated by the different photometric redshift codes. A number significantly lower than 68\% in the 68.3\% column indicates that errors are underestimated, and vice versa.
}
\label{table5}
\end{table*}
We find that a majority of codes return underestimated confidence intervals, i.e., fewer than $\sim$68\% and 95\% of the galaxies with known spectroscopic redshifts fall within the estimated error intervals of the photometric redshifts. There are two main factors affecting the derived $\chi^2$~values, P(z) distributions, and widths of the derived 68\% and 95\% intervals. First, the size of the quoted photometric errors in the photometric redshift fitting may affect results in the sense that systematically underestimated errors may drive $\chi^2$~to high values and result in narrow P(z) distributions. On the other hand, photometric errors that are unrealistically large decrease the $\chi^2$~values. This could result in seemingly acceptable fits over a larger redshift range and therefore a broad P(z) distribution and an overestimate of the confidence intervals. A difference between the codes compared here is that some have added extra smoothing errors to existing photometric errors (codes shown in Table \ref{table1}). Adding extra errors will effectively work as a smoothing of the P(z) distributions and result in relatively larger numbers in Table \ref{table5} compared to what the original photometric errors would result in. For example, codes 3B and 12I, which have the largest fractions quoted, are among the codes adding the largest smoothing errors to the existing photometric errors. Secondly, the completeness of the template SED set used affects derived $\chi^2$~values and associated P(z) distributions. Utilizing a coarse set of templates that does not sufficiently cover the true SED distribution, may result in acceptable $\chi^2$~value from only at a very narrow range of redshifts. This could lead to a narrow probability distribution and an underestimate of the confidence intervals. In Table \ref{table5}, the small values for code 5D is likely due to a relatively coarse grid of template SEDs. Therefore, even if the photometric redshifts agree well with the spectroscopic control sample, one should be cautious when using the errors for photometric redshifts if these are based on the results from the $\chi^2$~fitting. In Section 5.2, we describe a simple method for adjusting the quoted errors so that they better reflect the actual uncertainty suggested by the spectroscopic control sample.

\subsection{Closer look at outliers}
Table \ref{table2} shows that the outlier fraction for the $H$-band selected catalog lies in the range $\sim$4-15\%, depending on code. When comparing only the five codes with the lowest scatter, the range of outliers is narrowed to 3.6-5.3\%. In absolute numbers, this corresponds to 21-31 objects per code of the total 589 objects in the spectroscopic control sample. The number of individual objects flagged as an outlier by at least one of the five codes is 48. Of these, 20 are flagged by one code only, 7 by two codes, 2 by three codes, 8 by four codes, and 11 by all five codes. If we look at the case with the median photometric redshift from the five codes with the lowest scatter, we find an outlier fraction of 3.1\%, corresponding to 18 objects. Of these objects, 7 and 11 are flagged as outliers in 4 and 5 codes, respectively. The fact that 18 outliers are flagged by at least 4 of the 5 codes indicates that some feature drives the photometric redshift to an outlier independent of code or template SED used. These objects may have an SED not represented by any of the template SED sets. Otherwise, the spectroscopic redshift could be incorrect or there could be problems with the photometry. To investigate this, we look closer at the spectra for the subsample of 18 objects flagged as outliers by the median method. We find that at least 12 objects have spectroscopic redshifts that most likely are not the highest quality and could therefore be wrong. There are objects with spectra measured by different groups that disagree. A few of the objects also have close companions (within $\sim$1 arcsec) where it is difficult to determine if the correct object in the photometric catalog has been assigned the spectroscopic redshift. So it is possible that the actual outlier fraction for the combined median photometric redshift is significantly less than reported in Table \ref{table2} and Table \ref{table3}, perhaps as low as $\sim$1\% when using the median method.

\section{Combining results to improve photometric redshifts}
We have shown that combining results from multiple codes leads to photometric redshifts with lower scatter and outlier fraction than any individual code. This important result implies that using a combination of outputs from multiple algorithms can significantly improve the quality of photometric redshifts. The fact that the median outperforms any individual method indicates that net systematic errors must go in opposite directions amongst different codes, such that the middle value will have smaller scatter about the true redshift than even the best single technique. We expect systematic errors to vary due to differences in the templates used, priors applied, or fitting algorithms employed. In effect, there is a 'wisdom of crowds' in combining results from different photometric redshift codes, much like can occur when combining multiple estimates of quantities in other fields (Surowiecki 2005).

Besides deriving accurate photometric redshifts, we are also interested in assigning proper errors to derived photometric redshifts. In this section, we look more in detail into these issues by investigating different ways of combining data when we have results derived independently by different participants. For this particular investigation, we use results from codes number 3B, 6E, 7C, 11H, and 13C. For each code, we have the calculated photometric redshift and the full redshift probability distribution, P(z), tabulated in the range $0<z<7$~in steps of $\Delta z=0.01$. Different codes use different recipes for assigning the photometric redshift based on the P(z). Either the highest peak can be used to determine the photometric redshift, or some kind of weighted photometric redshift can be derived by integrating over the probability distribution. To get a clean comparison between methods, we use below photometric redshifts based on both the peak of the P(z), i.e., $z_{peak}$, as well as the weighted photometric redshift, $z_{weight}$, and compare results separately. We compute the latter by integrating over the main peak of the P(z) distribution. We do not want to integrate over the full P(z) distribution since there are cases with multiple peaks due to e.g., the aliasing between the Lyman and the 4000\AA~breaks (where the actual P(z) could be basically zero at the reported photometric redshift if it falls between two peaks).

\subsection{Method 1: Straight median}
As already shown above, if we compare the median photometric redshift from multiple codes for each individual object with the spectroscopic control sample, we get a scatter and an outlier fraction lower than any individual code. The resulting scatter and outlier fraction from the straight median is shown in the first two rows of Table \ref{table6}. These results indicate that combining results from multiple codes is advantageous. However, using a strict median does not directly produce any useful photometric redshift error estimate. Basing the errors on the scatter between the five codes will not yield a consistent measurement because of the expected highly non-Gaussian shape of the photometric redshift P(z) and the strong possibility that the various photometric redshift estimates are covariant with each other (e.g., they are based on the same photometry), so their scatter will not reflect all measurement uncertainties. We therefore look into a few more ways of combining data that may provide accurate results for both the photometric redshifts and the errors. There is no significant difference between using $z_{peak}$~compared to  $z_{weight}$.

\subsection{Method 2: Adding probability distributions}
As a second approach we add the full P(z)$_i$~distributions from the different codes to produce a combined P(z). From Table \ref{table5} we saw that a number of codes underestimate the errors, i.e., the distributions are too peaked around the derived photometric redshift. This will bias the combined redshift towards the values given by codes that underestimate the errors. At the same time, the photometric redshift of codes that overestimate the error will be given lower weights. To alleviate this, for codes underestimating the errors, we smooth each P(z)$_i$~using a simple recipe where we for each redshift bin $j$~replace the probability with a combination of three adjacent bins P(z$_j$)$_i$=0.25P(z$_{j-1}$)$_i$+0.5P(z$_j$)$_i$+0.25P(z$_{j+1}$)$_i$. We recalculate the fraction of the spectroscopic sample inside the 68.3\% interval and iterate this procedure until the correct fraction is recovered. We thereafter apply the same smoothing, individually calculated for each code, to the full sample of galaxies. For the codes that overestimate the errors, we instead use a simple model to sharpen the P(z)$_i$. For each code we set P(z$_j$)$_i$=P(z$_j$)$_i^{1/\alpha}$, adjusting the exponent $\alpha$~so the correct 68.3\% of the galaxies in the spectroscopic control sample falls inside the 68.3\% confidence interval. After normalizing each P(z)$_i$ to unity, we add all five distributions and renormalize. 

\begin{figure*}
\epsscale{1.3}
\plotone{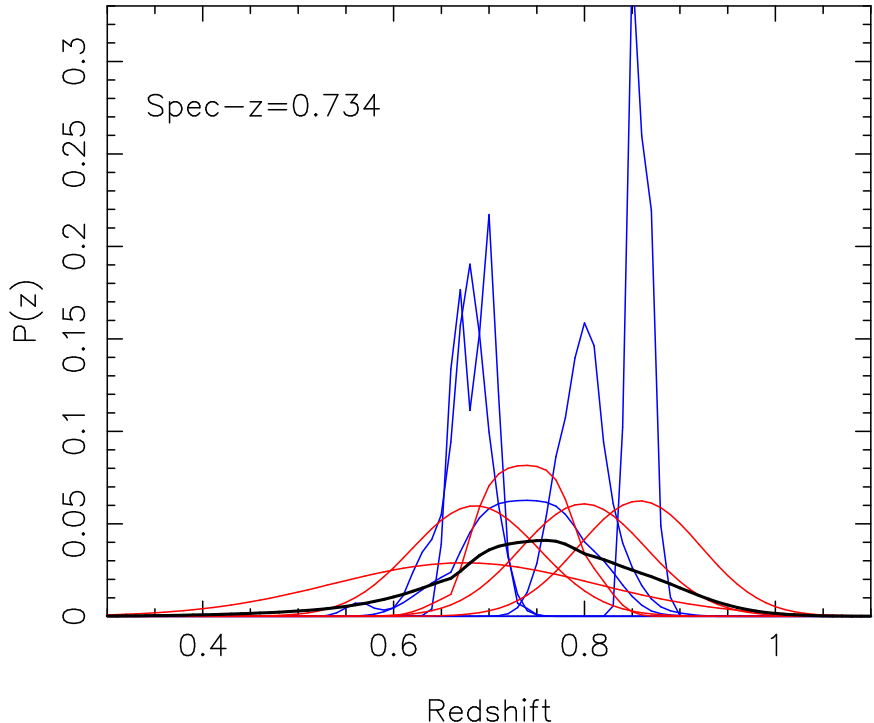}
\figcaption[f14.eps]{An example of the photometric redshift probability distributions for one galaxy with spectroscopic redshift $z$=0.734. Blue lines show five individual codes (code 3B, 6E ,7C ,11H, and 13C) without correcting distributions so that they match the 68.3\% confidence interval criterion. Red lines show the distributions after corrections. Finally, the black line shows the sum of the individual distributions.
\label{fig14}}
\end{figure*}

To illustrate this procedure, we show in Figure \ref{fig14} an example applied to a galaxy with spectroscopic redshift $z$=0.734. The five blue lines show the probability distributions for five individual codes (codes 3B, 6E, 7C, 11H, and 13C). To account for the four codes underestimating the error intervals and one code overestimating them, we apply the smoothing and sharpening described above. This should lead to distributions with more consistent confidence intervals. The resulting individual distributions are shown with red curves in the figure. In this particular case, there is one code that produces a P(z) with a double peak, which turns into a single peak after smoothing. After adding the five individual distributions, the resultant distribution is shown with the black line.

In Table \ref{table6}, we show the results from adding the probability distributions in rows three and four. Compared to the straight median, the combined P(z) results in slightly higher outlier fraction and $\sigma_F$, but similar $\sigma_O$. Therefore, either method should result in photometric redshifts with no significant difference in accuracy. The advantage with the added P(z)$_i$~method is that it provides an estimate of the full probability distribution, which could be used to calculate e.g., 68.3\% confidence intervals. To test how well the combined P(z) distributions reflects the true errors, we repeat the exercise above and calculate the fraction of objects in the control sample that falls within the 68.3\% interval of the combined P(z). We find that 85\% of the spectroscopically determined redshifts fall within the 68.3\% confidence intervals. This suggests that combining the P(z) by adding the individual distributions overestimates the size of the 68.3\% confidence intervals. To get a distribution that better represent the errors, we sharpen the distribution to recover 68.3\% of the control sample within the 68.3\% confidence interval, as described above.

\subsection{Method 3: Hierarchical Bayesian Approach}
As an alternative to a straight addition of the probability distributions, we adopt a hierarchical Bayesian approach following the method in Lang \& Hogg (2012) (similar methods were employed by Press (1997) and Newman et al. (1999)). We want to determine the consensus P(z) for each object accounting for the measured probability distributions (hereafter P$_m$(z)$_i$) may be wrong. We call the fraction of measurements that are bad $f_{\rm bad}$~and write for each code $i$
\begin{equation}
P(z,f_{\rm bad})_i = P(z | {\rm measurement~is~bad})_i f_{\rm bad} +
\end{equation}
$ P(z | {\rm measurement~is~good})_i(1-f_{\rm bad})$.\\
\vskip-0.1truecm
Here P(z $\vert$~measurement is bad) (hereafter U(z)) is a redshift probability distribution that we assume in the case where the observed P$_m$(z)$_i$~is wrong. We assume that there is no information on the redshift if the measurement is bad and therefore set U(z) to be uniform for all different codes. For the redshift range $0<z<7$ used, this means U(z)=1/7. We now have
\begin{equation}
P(z,f_{\rm bad})_i = \frac{1}{7} f_{\rm bad} + P_m(z)_i(1-f_{\rm bad}).
\end{equation}
The combined $P(z,f_{\rm bad})$~for all five measurements can be calculated as
\begin{equation}
P(z,f_{\rm bad})=\prod_{i=1}^5 P(z,f_{\rm bad})_i ^{1/\alpha}.
\end{equation}
Here $\alpha$~is a constant reflecting the degree of covariance between the results from the different codes (see below). We finally marginalize over f$_{\rm bad}$~to get the redshift probability distribution for each object
\begin{equation}
P(z)=\int_0^1 P(z,f_{\rm bad})df_{\rm bad}
\end{equation}
From the resulting P(z) we can determine the photometric redshift as either the peak of the distribution, $z_{peak}$, or the integral of the main feature in the distribution,  $z_{weight}$. In Table \ref{table6}, we show the resulting scatter between the photometric redshifts and the spectroscopic control sample. Similar to the methods described in Section 5.1 and 5.2, the Bayesian method produces a scatter that is lower than any of the individual codes. Compared to the straight median and the combined P(z) method, there is no significant difference. 

\begin{figure*}
\epsscale{1.3}
\plotone{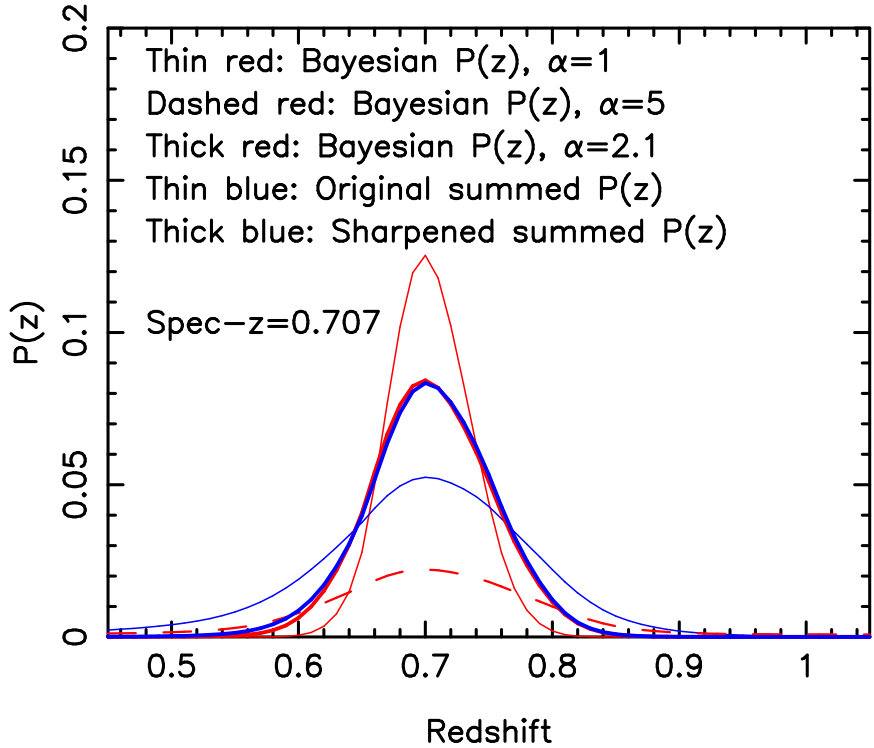}
\figcaption[f15.eps]{An example of the photometric redshift probability distributions for one galaxy with spectroscopic redshift $z$=0.707 derived using the Bayesian method with $\alpha$=1 (thin red line) and $\alpha$=5 (dashed red line) as well as after a straight summation of the individual distributions (thin blue lines). Thick red line shows the distribution for the Bayesian method when using $\alpha$=2.1, the value that recovers the correct 68\% of the spectroscopic control sample within the 68\% confidence interval. Finally, the thick blue line shows the result after having sharpened the distribution resulting from the summation method so that this also produces consistent 68\% confidence intervals.. 
\label{fig15}}
\end{figure*}

In Equation (5), $\alpha$~can adjust for any covariance between the different individual results. Setting $\alpha$=1 is equivalent to assuming statistical independence between all codes, while setting $\alpha$=5, i.e., the number of codes that are combined, corresponds to assuming full covariance. In this case, we expect some degree of covariance, both because all the photometric redshift estimates are based on identical photometry, and because there are overlaps between the five codes in templates and methods. The peak redshift  of the resulting photometric redshift does not depend on the value of $\alpha$; however, the width of the final P(z) distribution does. We find that using $\alpha$=1 underestimates the errors; only 46\% of the  objects in the spectroscopic control sample fall inside the calculated 68\% confidence interval. On the other hand, setting $\alpha$=5 overestimates the errors; 91\% of the objects in the spectroscopic control sample fall inside the 68\% confidence interval. To make the resulting P(z) distributions consistent with the spectroscopic control sample, we derive the value of  $\alpha$~that recovers 68\% of the spectroscopic redshifts within the 68\% confidence intervals of the derived P(z) distributions. This is achieved for $\alpha$=2.1. Ignoring the impact of priors and $f_{\rm bad}$, setting $\alpha$=5 would be equivalent to averaging the predicted $\chi^2(z)$~curves from each code, as opposed to averaging the $P(z)$~estimates as in Section 5.2. Figure \ref{fig15} shows the output P(z) of a single object for a number of cases, as an example the effect the choice of $\alpha$~has on the Bayesian method and sharpening of P(z) distributions in the summation method. For the Bayesian method, we show the results with $\alpha$=1 (thin red line), $\alpha$=5 (dashed red line) and $\alpha$=2.1 (thick red line). It is clear that lower $\alpha$~produces narrower P(z) distributions. The result from the straight summation is shown with the thin blue line, while the result after sharpening the P(z) distribution so that the control sample recovers the expected 68\% of the galaxies within the 68\% confidence interval is shown with the thick blue line. Although the final P(z) distributions for the two methods are derived using completely different algorithms, they produce very similar results. Note that $\alpha$ ~and the sharpening are not calculated particularly for this object, but are derived as averages for the full control sample. 

Inspecting the bias$_z$~values in Table \ref{table6}~shows that the shift is small for all methods, mean[$\Delta z/(1+z_{spec})$]$<$0.01. The uncertainty in the bias values are typically $\sigma_{bias_z}\le 0.003$, indicating that the bias is statistically non-zero at a $\sim$3$\sigma$~level. However, in every case the bias is significantly smaller than the scatter, so the latter will dominate the statistical risk.

The similarities between the results suggests that either the Bayesian method or the straight adding of the P(z) distributions (after sharpening or smoothing the individual P(z)) could be used to derive the photometric redshifts and probability distributions.

In this example of the hierarchical Bayesian method, we have used a simple assumption for U(z), i.e., that we have no information if the measured P$_m$(z)$_i$~is wrong. Furthermore, we have allowed $ f_{\rm bad}$~in the whole range $ f_{\rm bad}$=[0.0,1.0]. Alternatively, we can assume that there is at least some minimum probability that the actual measurement are correct and let the bad fraction vary in the range $ f_{\rm bad}$=[0.0,x]. Repeating our analysis after varying x does not change results significantly, however, there is a slight decrease in the outlier fraction and full rms when setting $0.3<x<0.5$, i.e., assuming that the measured P(z)$_i$~are correct at least 50-70\% of the times. Setting x=0.0, equivalent to assuming that all measured P(z)$_i$~are always correct, does, however, result in a significant increase in the outlier fraction (from 3.4\% to 4.9\%) and full rms ($\sigma_F$=0.10 to $\sigma_F$=0.36). 

The example above illustrates that the hierarchical Bayesian approach does indeed provide means for improving results. It is possible to assume a more advanced guess for the shape of  U(z). For example, if the measurement is bad, one could use a redshift probability following the volume element redshift dependence. Using this assumption, we find that the outlier fraction slightly decreases (from ~3.4\% to ~3.1\%), while the full rms show a marginal increase ( $\sigma_F$=0.10 to  $\sigma_F$=0.11) and (after excluding outliers) the rms, $\sigma_O$, remains unchanged. Since we do not expect the spectroscopic control sample to follow the distribution of the volume element, we do not expect this example necessarily reflects the true expected effect of the volume element assumption. 

A further refinement of the model would be to assume that the redshift distribution of a bad measurement follows the expectations of an assumed luminosity function combined with a magnitude limit appropriate for this particular survey. In addition, it should be possible to let the expected distribution be dependent on, e.g., apparent magnitude or color. 

Instead of using a generic form for U(z), another possibility is to dilate the given P(z) and use this for U(z). In this case we assume that the errors are underestimated if the measurement is bad, rather than having no information. There are many possibilities when applying the hierarchical Bayesian method as discussed in Lang \& Hogg (2012).

\begin{table*}
\caption{Photometric redshift accuracy when combining results from multiple codes}
\centering
\begin{tabular}{lccccccc}
Method & bias$_z$ & OLF & $\sigma_F$ & $\sigma_O$ & $\sigma_{NMAD}$ &  $\sigma_{dyn}^f$ & OLF$_{dyn}^g$   \\
\tableline 
\tableline
Straight median of $z_{peak}$ & -0.009 & 0.031 & 0.078 & 0.0296 & 0.025 & 0.024 & 0.056\\
Straight median of $z_{weight}$ & -0.008 & 0.031 & 0.079 & 0.0296 & 0.025 & 0.024 & 0.056\\
\tableline
Combined P(z), using $z_{peak}$ & -0.006 & 0.044 & 0.108 & 0.0293 & 0.024 & 0.025 & 0.066\\
Combined P(z), using $z_{weight}$ & -0.010 & 0.041 & 0.105 & 0.0303 & 0.029 & 0.026 & 0.060\\
\tableline
Bayesian using $z_{peak}$ & -0.007 & 0.034 & 0.099 & 0.0299 & 0.025 & 0.025 & 0.061\\
Bayesian using $z_{weight}$ & -0.007 & 0.034 & 0.098 & 0.0296 & 0.026 & 0.025 & 0.058\\
\tableline
\end{tabular}
\tablecomments{
Table shows photometric redshift accuracy using different method for combining results from five separate codes (code 3B, 6E, 7C, 11H, and 13C). Taking a straight median of the five is shown on top. In the middle, results are shown after adding the full redshift probability distributions for each code. Bottom results show the accuracy after using a hierarchical Bayesian method when combining distributions. For each case we show the results after adopting both the peak of the probability distribution ($z_{peak}$) and the weighted mean of the distribution ($z_{weight}$) as the photometric redshift. See Table 2 for the definition of columns 2 to 8.
}
\label{table6}
\end{table*}

\section{Comparison to earlier work}
Over the years, there has been a number of investigations comparing results from different codes in order to assess the accuracy of and the consistency between different photometric redshift codes. This includes Hogg et al. (1998), Abdalla et al. (2008),  and Hildebrandt et al. (2008, 2010).  The most comprehensive previous investigation of photometric redshift methods conducted in a similar way to what presented here is described in Hildebrandt et al (2010). In that investigation, the result of twelve different runs, representing eleven codes, are presented. Of these codes, three are common to this investigation (EAZY, LePhare, and HyperZ). Photometric redshifts are calculated using an $R$-filter selected 18-band photometry catalog covering the GOODS-North field. The wavelength range covered is the same as here, i.e., U-band to the IRAC 8.0$\mu$m channel. The spectroscopic sample includes $\sim$2000 objects, of which one quarter was provided as a training sample. The overall scatter after excluding outliers lies in the range $\sigma_O$=0.04-0.08, with a median of the twelve runs of $\sigma_O$=0.059. This is slightly higher than the median found here  $\sigma_O$=0.046 (using the $z$-band selected results in Table \ref{table3}). More importantly, the outlier fraction in Hildebrandt et al. lies in the range 8-31\% and has a median of 18.5\%, while our investigation reports outlier fractions 4-14\% with a median 6.4\%. This significant difference, despite the many similarities in setup, could be due to a number a reasons. We have here used a uniformly produced photometry over the whole wavelength range using the TFIT method, while Hildebrandt et al. used coordinate matching between three different data sets (ground-based optical/NIR, $HST$/ACS, and $Spitzer$/IRAC). This could introduce biases in the photometry due to blending, mismatches and differences in apertures used. Furthermore, we have made an effort to include only the highest quality spectroscopic redshifts and have excluded all known X-ray and radio sources when compiling our training and control samples.  This should assure us an unbiased estimate of the scatter and outlier fractions when comparing spectroscopic and photometric redshifts. At the same time, Hildebrandt et al. reports that at least some of the high outlier fraction could be due to X-ray sources or the spectroscopic sample used. We therefore think that the outlier fractions of a few per cent found in our study should be more representative of what is achievable with photometric redshifts when using deep high quality photometry.

\section{Conclusions and summary}
We have used the CANDELS  GOODS-S $HST$ ~WFC3 $H$-band and ACS $z$-band selected catalogs containing uniform TFIT photometry covering the $U$-band to IRAC infrared bands to investigate the behavior of photometric redshifts. Using a control sample with high quality spectroscopic redshifts, we have compared photometric redshifts derived from a number of different codes. We have investigated how the accuracy of the photometric redshifts depends on code and template SED set used. We have also investigated the dependence on redshift, galaxy color and brightness. Finally, we discussed combining results from multiple codes for improving the photometric redshifts and deriving reliable error estimates. 
Our main conclusions are 
\begin{itemize}   

\item{There is no particular code or template SED set that produces significantly better photometric redshifts compared to others. However, the codes that produce the best photometric redshifts all include training using a spectroscopic sample to calculate offsets or shifts to either the photometric zero-points or the template SEDs.}

\item{There is a strong magnitude dependence on the accuracy of the photometric redshifts: rms values calculated for a spectroscopic control sample are only valid at the magnitudes probed by that sample. The photometric redshift uncertainty is likely to be significantly larger for a catalog that is deeper than the spectroscopic subsample.} 

\item{We investigated the redshift dependence of the scatter between photometric redshifts and a control sample of spectroscopic redshifts and find that the rms, when normalized to redshift by $\sigma$=rms$[(z_{phot}-z_{spec})/(1+z_{spec})]$, is almost independent of redshift. On the other hand, the fraction of outliers is elevated in the range $2.2<z<3.7$, possibly due to the relatively weak Lyman break signal in the lower part of this range, as well as aliasing between the Lyman and the 4000\AA~breaks. The outlier fraction at high redshift ($z>3.7$) is low due to the strong Lyman break signal.}

\item{We find that the rms is only weakly dependent on galaxy color as measured by the rest frame $B-V$~color. Only for the very reddest early-type galaxies is there an indication that the scatter is smaller than the rest of the galaxy population. There is no increase in scatter for the most blue galaxies that should have the smallest 4000\AA~breaks.}

\item{The bias$_z$~between the photometric and spectroscopic redshifts, defined as mean[$(z_{spec}-z_{phot})/(1+z_{spec})$] after excluding outliers is statistically inconsistent with zero at a significance of $\gsim 3 \sigma$. However, the bias is always smaller than the scatter and the latter therefore dominates the total uncertainty.}

\item{The photometric redshift codes produce an estimate of the uncertainty in the derived photometric redshift either as a full redshift probability distribution, P(z), or as quoted confidence intervals corresponding to e.g., 68.3\% or 95,4\% confidence intervals. Using the spectroscopic control sample with known redshifts, we calculate which fraction of the galaxies falls inside the 68.3\% or 95.4\% confidence intervals for the different codes. We find that a majority of the codes produce confidence intervals that are too narrow compared to expectations, i.e., the errors in the photometric redshifts are most often underestimated. Factors contributing to the narrow distributions could be underestimated photometric errors or too coarse set of template SEDs. We describe a method for adjusting probability distributions so that the correct fraction of galaxies in the control sample falls inside a specified confidence interval.}

\item{We can derive photo-z with lower scatter and outlier fraction when we combine results from different codes, when compared to any single code. Taking a straight median, using a sum of the individual probability distributions, or using a hierarchical Bayesian method yields very similar results. The two latter methods produce a probability distribution that can be used to assign errors to the photometric redshifts. For our spectroscopic sample, we find an rms of $\sigma_O \sim 0.03$~with an outlier fraction of at most $\sim$3\%.}

\end{itemize}

We finally note that the photometric redshifts presented here are based on test catalogs derived from a subset of CANDELS GOODS-S data. After including additional data, particularly the full depth $HST$/WFC3 $J$- and $H$-bands, we expect further improvements in the absolute values of the photometric redshift accuracies. Further improvements are possible by the addition of medium and narrow band data that are available for the CANDELS fields. The CANDELS GOODS-S photometric redshift catalog will be made publicly available and is described in T. Dahlen et al. 2013 (in prep.).

\acknowledgments{
We are grateful to our referee, Giuseppe Longo, for providing valuable comments and suggestions for improving this paper.
Based on observations made with the NASA/ESA Hubble Space Telescope, obtained at the Space Telescope Science Institute, which is operated by the Association of Universities for Research in Astronomy, Inc., under NASA contract NAS 5-26555. These observations are associated with programs GO-9352, GO-9425, GO-9583, GO-9728, GO-10189, GO-10339, GO-10340, GO-11359, GO-12060, and GO-12061. 
Observations have been carried out using the Very Large Telescope at the ESO Paranal Observatory under Program ID(s): LP168.A-0485.
This work is based in part on observations made with the {\it Spitzer Space Telescope}, which is operated by the Jet Propulsion Laboratory, California Institute of Technology, under a contract with NASA. Support for this work was provided by NASA through an award issued by JPL/Caltech.
}
%% If more than five, use three and et al.

\end{document}